\documentstyle[aps,eqsecnum,preprint,prd,epsfig]{revtex}
\begin{document}
\draft
\preprint{BARI-TH 225/96}
\title{ Perturbation Theory with a Variational Basis:\\
the Generalized Gaussian Effective Potential}
\author{Paolo Cea and Luigi Tedesco} 
\address{Dipartimento di Fisica and Sezione INFN \\
Via Amendola 173,  I-70126 Bari, Italy}
\maketitle
\begin{abstract}
The perturbation theory with a variational basis is constructed and analyzed. 
The generalized Gaussian effective potential is introduced and evaluated up to 
the second order for selfinteracting scalar fields in one and two spatial 
dimensions. The problem of the renormalization of the mass is discussed in 
details. Thermal corrections are incorporated. The comparison between the 
finite temperature generalized Gaussian effective potential and the 
finite temperature effective potential is critically analyzed. 
The phenomenon of the restoration at high temperature of the symmetry 
broken at zero temperature is discussed.
\end{abstract}
%
%
\section{INTRODUCTION}

In the recent years the variational Gaussian approximation has played an 
important role in the non perturbative study of quantum field theories. 
In particular, to investigate the spontaneous symmetry breaking phenomenon 
in scalar quantum field theories it has been introduced the Gaussian effective
 potential~\cite{Stev84}. The main disadvantage of the variational approach 
is the absence of the control of the approximation. Moreover, in quantum field 
theories the presence of ultraviolet divergences often makes useless the 
variational calculations.

The aim of this paper is to develop a variational scheme in scalar quantum field 
theories which allows to evaluate in a systematic manner the corrections to the 
Gaussian approximation and, at the same time, to keep under control the 
ultraviolet divergences. To this end, we shall construct a 
perturbation theory with a variational basis. The method we shall follow is
 widely used in many-body theory where it is known as the method of
correlated basis functions~\cite{Feenberg69}. Using the variational basis
 we construct a vacuum state  $|\Omega \rangle$ which is adiabatically
 connected to the Gaussian trial vacuum. Whereupon, we introduce the
 generalized Gaussian effective potential $V_G(\phi_0)$ defined as the 
 expectation value of
 the Hamiltonian density on $|\Omega \rangle$ in presence of the scalar
 condensate $\phi_0$. We shall give an explicit formula for $V_G(\phi_0)$
 which is similar to the usual perturbative expansion of the effective
 potential by means of the Feynman vacuum diagrams.
 Moreover, we shall show that the variational-perturbation theory
 developed in this paper offers a solution to the ultraviolet divergences
problem in the variational approaches which is analogous to the usual 
 perturbative renormalization theory. For the sake of simplicity,                   
 we perform the explicit calculations in the case of selfinteracting scalar fields
 in one and two spatial dimensions. Indeed, these theories are
super-renormalizable, so that we only need to renormalize the mass.
 In the second part of the paper we discuss the finite temperature corrections
 to the generalized Gaussian effective potential. Moreover we critically
compare our approach to the finite temperature effective potential and the
 Gaussian potential.  

The plane of the paper is as follow. In Section 2 we discuss the Gaussian 
approximation in scalar field theories and introduce the Gaussian effective 
potential. In Section 3 we set up the variational basis starting from the trial 
Gaussian vacuum wavefunctional. Section 4 is devoted to the perturbation 
theory with the variational basis.The generalized Gaussian effective potential 
is discussed in Section 5. The calculations of the second order corrections 
to the Gaussian effective potential are presented in Section 6 where we discuss 
in detail the mass renormalization. In  Section 7 we introduce the finite 
temperature generalized Gaussian effective potential and evaluate the lowest 
order thermal corrections. The second order thermal corrections are 
explicitly evaluated in Section 8. Our conclusions are draw in Section 9.  
Several technical details are relegated in two Appendices. In Appendix A 
we perform the high-temperature  expansions which are relevant for the 
lowest order thermal corrections. In Appendix B we collect some well-known 
result on the thermodynamic perturbation theory in the Matsubara's scheme. 
Moreover we present some useful results on the thermal propagator. 

\section{THE GAUSSIAN APPROXIMATION}

In this Section we discuss the Gaussian approximation in scalar quantum field 
theories. In particular we shall focus on the Gaussian effective potential 
\cite{Stev84} for selfinteracting scalar fields in $d=\nu+1$ space-time 
dimensions.

The Gaussian approximation in quantum field theories has been widely developed
since long time \cite{S-R,C-J-T,Barnes}. The Gaussian approximation is 
a variational method in which one considers trial Gaussian wavefunctionals as 
the ground state of the theory. 

Let us consider a real scalar field $\phi(x)$ whose Hamiltonian is 
\begin{equation}
\label{eq2.1}
H=\int d^{\nu}x \left[{\Pi^2(\vec{x}) \over 2} 
      + {1\over 2}\left(\vec{\nabla}
    \phi(\vec{x}) \right)^2 + {1\over 2} m^2 {\phi^2(\vec{x})} +
    {\lambda \over 4!} {\phi}^4(\vec{x}) \right] \, .
\end{equation}
In the Schr\"odinger representation the physical states are wavefunctional of 
$\phi$; the conjugate momentum $\Pi(x)$ acts as functional derivative 

\begin{equation}
\label{2.2}
\Pi(x) |\Psi\rangle  \,\,\, \longrightarrow \,\,\, {1\over i}
{\delta\over {\delta \phi(\vec{x})}} \,\, \Psi[{\phi}] \, .
\end{equation}
The inner product is defined by

\begin{equation}
\label{2.3}
<\Psi_1|\Psi_2> = \int [d \phi] \,\, \Psi_1^*[\phi] 
\,\, \Psi_2 [\phi] \, .
\end{equation}
The stationary Schr\"odinger equation reads
\begin{equation}
\label{eq2.4}
\int d^{\nu}x \left[-{1\over 2}{\delta^2\over 
    {\delta\phi(\vec{x})\delta\phi(\vec{x})}} + {1\over 2}\left(\vec{\nabla}
    \phi(\vec{x}) \right)^2 + {1\over 2} m^2 {\phi^2(\vec{x})} +
    {\lambda \over 4!} {\phi}^4(\vec{x}) \right] \Psi[\phi]=E \, \Psi[\phi] \, . 
\end{equation}
The analogy with ordinary quantum mechanics is evident. In particular, we would 
like to apply the variational principle which has been successfully developed in 
quantum mechanics.

In quantum field theories it is the ground state that determines the physical 
properties of the quantum system. The Gaussian approximation amounts to 
approximate the vacuum functional with a set of trial Gaussian functionals 
centered at $\phi_0$:

\begin{equation}
\label{eq2.5}
\Psi_0[\phi]={\cal N} \, \exp\left[-{1\over 4}\int d^{\nu}x \, d^{\nu}y \,
[ \phi(\vec{x})
- \phi_0 ] \,  G(\vec{x},\vec{y}) \, [ \phi(\vec{y}) - \phi_0 ] \,
\right]
\end{equation}
where the normalization constant is such that:
\begin{equation}
\label{eq2.6}
\langle \Psi_0|\Psi_0 \rangle=1 \, .
\end{equation}
A well-known method to investigate the structure of a quantum field theory 
is to use the effective potential $V_{eff}(\phi_0)$ \cite{Col}.
In scalar field theories it turns  out that the effective potential is the 
expectation value of the Hamiltonian density in a certain state wherein  the 
expectation value of the scalar field is $\phi_0$ \cite{J-L}. These 
considerations suggested to introduce the so-called Gaussian effective
potential $V_G(\phi_0)$. 

The Gaussian effective potential is defined by minimizing the Hamiltonian 
density on the set of wavefunctionals Eq.(\ref{eq2.5}):

\begin{equation}
\label{eq2.7}
V_{GEP}(\phi_0)={1\over V} \min_{| \Psi_0 \rangle} \langle 
\Psi_0|H|\Psi_0\rangle
\end{equation}
where $V$ is the spatial volume.

$V_{GEP}(\phi_0)$, being a variational quantity, not only goes  beyond the 
perturbation theory, but often gives a more realistic picture of the 
qualitative physics than the effective potential. 
Moreover the Gaussian effective potential is easily computable. 
To see this, we note that
due to the translation invariance of the vacuum we have:

\begin{equation}
\label{eq2.8}
G(\vec{x}, \vec{y}) = \int {d^{\nu}k\over {(2 \pi)^{\nu}}} \, 
e^{i\vec {k} \cdot (\vec {x} - \vec {y})} \, 2 \, g(\vec{k}) \, .
\end{equation}
Let us consider the following functional 

\begin{equation}
\label{eq2.9}
\Psi_0^J[\phi]={\cal {N}}exp \left[ -{1\over 4}\int d^{\nu} x \,d^{\nu} y \,
\eta(\vec{x}) \, G(\vec{x}-\vec{y})\, \eta(\vec{y}) +
{1\over 2}\int d^{\nu}  x\, \eta(\vec{x}) J(\vec{x}) \right] \, ,
\end{equation}
where

\begin{equation}
\label{eq2.10}
\eta(\vec{x})=\phi(\vec{x})-\phi_0 \, ,
\end{equation}
and  ${\cal N}$ is fixed by Eq. (\ref{eq2.6}). We can easily evaluate the following 
functional 

\begin{equation}
\label{eq2.11}
I[J]=\langle \Psi_0^J | \Psi_0^J \rangle \, .
\end{equation}
Indeed, Eq. (\ref{eq2.11}) involves a straightforward Gaussian functional 
integration. We get 

\begin{equation}
\label{eq2.12}
I[J]=exp\left[{1\over 2} 
\int d^{\nu} x \, d^{\nu} y \, J(\vec{x}) \, G^{-1}(\vec{x}-\vec{y})
\, J(\vec{y})\right]
\end{equation}
where

\begin{equation}
\label{eq2.13}
G^{-1}(\vec{x},\vec{y})=\int{{d^{\nu}k}\over {(2 \pi)^{\nu}}} {1\over {2 g(
\vec{k}})} e^{- i \vec{k} (\vec{x}-\vec{y})}.
\end{equation}
Now, we have:

\begin{equation}
\label{eq2.14}
\langle \Psi_0| \eta(\vec{x_1})...\eta(\vec{x_n}) | \Psi_0 \rangle = \left. 
{{\delta^n I[J]}\over {\delta J(\vec{x_1})...\delta J(\vec{x_n})}}
\right|_{J=0}.
\end{equation}
Equations (\ref{eq2.14}) and (\ref{eq2.12}) allow to evaluate the expectation 
values of monomial in $\eta(\vec{x})$ on the Gaussian vacuum functionals. 
It is now a straightforward exercise to calculate

\begin{equation}
\label{eq2.15}
E_0[\phi_0,g(\vec{k})]={{\langle \Psi_0|H|\Psi_0  \rangle}\over 
{\langle \Psi_0| \Psi_0 \rangle}} \; .
\end{equation}
We get 

\begin{eqnarray}
\label{eq2.16}
E_0[\phi_0, g(\vec{k})] &=& \;  V   \left \{
  {1\over 4} \int \frac {d^{\nu} k} {(2 \pi)
^{\nu}} g(\vec{k}) + \frac {m^2} {2} \phi_0^2 \frac {\lambda} {4!} \phi_0^4 +
{1\over 4} \int \frac {{\vec{k}}^2 + m^2 + {{\lambda}\over 2} \phi_0^2} 
{g(\vec{k})} \right. \nonumber \\
  && \left . + \frac {3}{4} \; \frac {\lambda} {4!} \; \left[\int  
 \frac {d^{\nu} k}
 {(2 \pi)^{\nu}} {1\over {g(\vec{k})}}\right]^2   \right\} \, .
\end{eqnarray}
The Gaussian effective potential is obtained by minimizing $E_0[\phi_0,g(
\vec{k})]$ with respect to $g(\vec{k})$. By imposing the extremum condition

\begin{equation}
\label{eq2.17}
{{\delta E_0[\phi_0,g(\vec{k})]}\over {\delta g(\vec{k})}}=0
\end{equation}
we obtain

\begin{equation}
\label{eq2.18}
g(\vec{k})=\sqrt{{\vec{k}}^2+\mu^2(\phi_0)} \; ,
\end{equation}
where $\mu(\phi)$ satisfies the gap-equation:

\begin{equation}
\label{eq2.19}
\mu^2=m^2 + {\lambda\over 2} \phi_0^2 + {\lambda\over 4} 
\int {{d^\nu k} \over {(2 \pi)^\nu}}{1\over {g(\vec {k})}} \, \; .
\end{equation}
Inserting Eqs. (\ref{eq2.18}) and (\ref{eq2.19}) into Eq. (\ref{eq2.16})
we get the Gaussian effective potential:

\begin{equation}
\label{eq2.20}
V_{GEP}(\phi_0)={\lambda\over {4!}}\phi_0^4 + {m^2\over 2}\phi_0^2 +
{1\over 2} \int {{d^\nu k} \over {(2 \pi)^\nu}} g(\vec{k}) - 
{\lambda\over {32}} 
\left[\int {{d^\nu k} \over {(2 \pi)^\nu}}{1\over {g(\vec {k})}} 
\right]^2.
\end{equation}
Putting \cite{Stev84}

\begin{equation}
\label{eq2.21}
I_n(\mu)={1\over 2} \int {{d^\nu k} \over {(2 \pi)^\nu}}
\, [{g(\vec{k})}^2]^{n-{1\over 2}} \, ,
\end{equation}
we rewrite Eq. (\ref{eq2.20}) in the more compact form:

\begin{equation}
\label{eq2.22}
V_{GEP}(\phi_0)={\lambda\over {4!}} \phi_0^4 + {m^2\over 2} \phi_0^2 +
I_1(\mu) - {\lambda\over {8}} I_0^2(\mu) \, ,
\end{equation}
while the gap  equation becomes 

\begin{equation}
\label{eq2.23}
\mu^2=m^2+{{\lambda}\over 2}\phi_0^2 -{{\lambda}\over 2} I_0(\mu) \; .
\end{equation}
For later convenience, it is useful to work in units of $\mu_0=\mu(\phi_0=0).$
Thus we introduce the following dimensionless parameters:

\begin{equation}
\label{eq2.24}
\Phi_0^2={{\phi_0^2}\over {\mu_0^{\nu - 1}}} \; ,
\end{equation}

\begin{equation}
\label{eq2.25}
\hat {\lambda}={\lambda\over {4!} \, \mu_0^{3-\nu}} \; ,
\end{equation}

\begin{equation}
\label{eq2.26}
x={\mu^2\over {\mu_0^2}} \; .
\end{equation}
Moreover we redefine the zero of the energy scale by subtracting in 
$V_{GEP}(\phi_0)$ the (divergent) quantity $V_{GEP}(\phi_0=0)$:

\begin{equation}
\label{eq2.27}
V^{\nu + 1}_{GEP}(\phi_0)= {{V_{GEP}(\phi_0) - V_{GEP}(\phi_0=0)}\over
{\mu_0^{\nu + 1}}} \; .
\end{equation}

\section{THE VARIATIONAL BASIS}

In the previous Section we have introduced the Gaussian effective potential.
The most serious problem of the Gaussian effective potential
resides in the lack of control on the variational approximation. For these 
reasons it is desirable to deal with a generalization of the Gaussian 
effective potential which allows to compute in a systematic way the corrections
to the Gaussian approximation \cite{Cea90}. The problem we are interested in is 
not an academic one. Indeed, it is well known that the Gaussian approximation
does not keep into account all the two loop contributions. As a consequence
in non abelian gauge theories the Gaussian approximation breaks the gauge 
invariance \cite{Cea88}.

In order to evaluate the corrections to the variational Gaussian approximation 
we need to set up a variational-perturbation theory. To this end we, now, 
construct a variational basis starting from the vacuum wavefunctional
$\Psi_0[\eta]$. To do this we consider $\Psi_0[\eta]$ as the ground state 
wavefunctional of a suitable Hamiltonian.

Let us consider  the following operators:

\begin{equation}
\label{eq3.1}
a(\vec {p})=\int {{d^{\nu} x}\over {(2 \pi)^{\nu\over 2}}}
{e^{ -i \vec{p}\cdot \vec{x}}\over {\sqrt{2 g(\vec{p})}}} 
\left(\int d^{\nu} y {1\over 2} G(x,y) \eta(y) +
{\delta\over {\delta \eta(\vec{x})}} \right)
\end{equation}

\begin{equation}
\label{eq3.2}
a^{^{\dag}}(\vec {p})=\int {{d^{\nu} x}\over {(2 \pi)^{\nu\over 2}}}
{e^{ i \vec{p}\cdot \vec{x}}\over {\sqrt{2 g(\vec{p})}}} 
\left(\int d^{\nu} y {1\over 2} G(x,y) \eta(y) -
{\delta\over {\delta \eta(\vec{x})}} \right).
\end{equation}
It is easy to see that the only non-trivial commutator is:

\begin{equation}
\label{eq3.3}
[a(\vec{p_1}), a^{\dag}(\vec{p_2})]=\delta(\vec{p_1}-\vec{p_2})\, .
\end{equation}
Moreover we have:

\begin{equation}
\label{eq3.4}
a(\vec{p}) \Psi_0[\eta]=0 \; .
\end{equation}

Now we rewrite the annihilation and creation operators Eqs.
(\ref{eq3.1}) 
and (\ref{eq3.2}) by means of the Fourier transform of the fluctuation 
fields $\eta(\vec{x}):$

\begin{equation}
\label{eq3.5}
\eta (\vec{p})=\int { {d^{\nu} x} \over { (2 \pi)^{\nu \over 2} }}
 \; e^{ - i \vec{p} \cdot \vec{x}} \; \eta(\vec{x}) \; ,
\end{equation}

\begin{equation}
\label{eq3.6}
{{\delta}\over {\delta \eta(\vec{p})}}= \int 
{ {d^{\nu} x} \over {(2 \pi)^{\nu \over 2}} } \; e^{ i \vec{p} \cdot \vec{x}}
 \;
{{\delta}\over {\delta \eta(\vec{x})}} \; .
\end{equation}
We get 

\begin{equation}
\label{eq3.7}
a(\vec{p})=\sqrt{{{g(\vec{p})}\over 2}} \left(\eta(- \vec{p}) + 
{1\over {g(\vec{p})}} {{\delta}\over {\delta \eta (\vec{p})}}\right),
\end{equation}

\begin{equation}
\label{eq3.8}
a^{\dag} (\vec{p})=\sqrt{{{g(\vec{p})}\over 2}} \left(\eta(\vec{p}) - 
{1\over {g(\vec{p})}} {{\delta}\over {\delta \eta (-\vec{p})}}\right).
\end{equation}
Consider, now, the Hamiltonian

\begin{equation}
\label{3.9}
{\tilde{H}}_0={1\over 2} \int d^{\nu} p \left[- {{\delta}\over {\delta \eta 
(\vec{p})}}  {{\delta}\over {\delta \eta (-\vec{p})}} + g^2(\vec{p})
\eta(\vec{p}) \eta(-\vec{p}) \right],
\end{equation}
which can be rewritten as

\begin{equation}
\label{eq3.10}
{\tilde{H}}_0=\int d^{\nu} p \,\, g(\vec{p}) \,a^{\dag} (\vec{p}) \,\,
 a(\vec{p}) + E_0 \; ,
\end{equation}
where

\begin{equation}
\label{eq3.11}
E_0={1\over 2} \int {d^{\nu} x} \int {{d^{\nu} p}\over {(2 \pi)^{\nu}}} 
\, g(\vec{p}) \; .
\end{equation}
Let $|0 \rangle$ be the vacuum of $H_0$ in the abstract ket formalism.
From Eq. (\ref{eq3.4}) it follows

\begin{equation}
\label{3.12}
\langle \eta | 0 \rangle = \Psi_0[\eta] \; ,
\end{equation}
that is $\Psi_0[\eta]$ is the vacuum of ${\tilde{H}}_0$ in the Schr\"odinger
representation. Starting from $|0 \rangle$ we can set up the many 
particle states by acting on the vacuum with the creation operators $a^{\dag}(
\vec{p})$. In the Schr\"odinger representation we have for
instance

\begin{eqnarray}
\label{3.12 b}
\Psi_1[\eta]= \langle \eta | \vec{p} \rangle = 
\langle \eta | a^{\dag}(\vec{p})|0\rangle & = &
\int \frac{ d^{\nu} x}{ (2 \pi)^{\frac{\nu}{2}}}
 \frac{e^{ i \vec{p}\cdot \vec{x}}}{\sqrt{2 g(\vec{p})}} 
\left(\int d^{\nu} y {1\over 2} G(x,y) \eta(y) -
{\delta\over {\delta \eta(\vec{x})}} \right) \Psi_0[\eta] \nonumber \\
 & = & \int \frac{ d^{\nu}x d^{\nu}y}{ (2 \pi)^{\nu\over 2} } \;
 \frac{e^{i \vec{p} \cdot \vec{x}}}{\sqrt{2 g(\vec{p})}} 
 \; G(\vec{x}-\vec{y}) \; \eta(\vec{y})
 \; \Psi_0[\eta] \; .
\end{eqnarray}
Obviously we also have 

\begin{equation}
\label{3.14}
H_0| \vec{p} \rangle = [E_0+g(\vec{p})] |\vec{p} \rangle \; ,
\end{equation}
and

\begin{equation}
\label{eq3.15}
\langle {\vec{p}}_1 | {\vec{p}}_2 \rangle = 
\delta({\vec{p}}_1 - {\vec{p}}_2) \langle 0|0 \rangle \; .
\end{equation}
In this way we construct the orthonormal set (Fock basis) of wavefunctionals 
$\{ \Psi_n[\eta] \}$, where $\Psi_n[\eta]$ is obtained by applying $n$ times 
the creation operator on $\Psi_0[\eta]$.

It should be emphasized that the Fock basis is univocally determined by 
the  vacuum functional $\Psi_0[\eta]$. As we will discuss in the next Section,
the vacuum functional will be fixed with a variational procedure. 
For this reason
the Fock basis $\{\Psi_n(\eta)\}$ will be referred to as a variational basis.

\section{PERTURBATION THEORY WITH THE VARIATIONAL BASIS}

In this Section we use the variational basis to set up a perturbation theory 
for the ground state energy. To this end we split our Hamiltonian Eq.(\ref{eq2.1})
into two pieces

\begin{equation}
\label{eq4.1}
H \, = \, H_0 \, + \, H_I\; ,
\end{equation}
where $H_0$ will be the ${\emph free}$ Hamiltonian  and $H_I$ the ${\emph
perturbation}$.
We define $H_0$ and $H_I$ as follows:

\begin{equation}
\label{eq4.2}
(H_0)_{nm}= \langle n | H_0 | m \rangle = \delta_{nm} H_{nn} \, ,
\end{equation}

\begin{equation}
\label{eq4.3}
(H_I)_{nm}= \langle n | H_I | m \rangle =(1- \delta_{nm}) H_{nm} \, ,
\end{equation}
where

\begin{equation}
\label{eq4.4}
H_{nm}=\langle n\,\, |\,\, H\,\, |\,\,  m \rangle
=\int [d \eta] \,\,\Psi_n^*[\eta]\,\, H\left(-i {\delta \over 
{\delta \eta}},\eta \right) \,\, \Psi_m[\eta] \, .
\end{equation}
Equations (\ref{eq4.2}) and (\ref{eq4.3}) show that the perturbation $H_I$
is given by the off-diagonal elements of the full Hamiltonian $H$ with respect
to the variational basis.
If the wavefunctional $\Psi_0[\eta]$ is close to the true ground state of $H$,
then we expect that the $(H_I)_{mn}$ are small with respect to 
$(H_0)_{nm}$, i.e. $H_I$ is a genuine perturbation.

We recall that in Section 2 we fixed the wavefunctional $\Psi_0[\eta]$ by
minimizing
$E[\phi_0, g(\vec{k})]=H_{00}$ on the class of trial Gaussian functionals,
Eq. (\ref{eq2.5}). In this way we get  an optimized perturbation expansion.
Moreover, we stress that in our scheme it is unnecessary to start with a small 
parameter in $H$. Thus our method goes beyond the usual perturbation theory.

We now address ourselves in the determination of $H_0$ and $H_I$. We evaluate,
firstly, the diagonal elements of $H$ with respect to the Fock variational 
basis $\{ \Psi_n[\eta]\}$. To this end we rewrite the Hamiltonian 
(\ref{eq2.1}) in terms of the annihilation and creation operators (\ref{eq3.7})
and (\ref{eq3.8}).  A rather length but otherwise straightforward calculation 
shows that: 

\begin{equation}
\label{eq4.5}
H=H^{(0)}+H^{(1)}+H^{(2)}+H^{(3)}+H^{(4)}
\end{equation}
where

\begin{equation}
\label{eq4.6}
H^{(0)}=E[\phi_0,g(\vec{k})] \; ,
\end{equation}

\begin{equation}
\label{eq4.7}
H^{(1)}=\left[m^2 \phi_0 + {\lambda\over 6} \phi_0^3 
+ {{\lambda}\over 4}\phi_0 \int 
{ {d^{\nu} k} \over {(2 \pi)^{\nu}} } {1\over {g(\vec{k})}}
\right] \int d^{\nu} x :\eta(\vec{x}): \; ,
\end{equation}

\begin{eqnarray}
\label{eq4.8}
H^{(2)}&=& {1\over 2} \int d^{\nu} p \left[g(\vec{p})+ { { {\vec{p}}^2 + m^2
+ {\lambda\over 2} \phi_0^2 }\over {g(\vec{p})}} + 
{\lambda\over 4} \int { {d^{\nu} k}\over {(2 \pi)^{\nu}}} 
{1\over { g(\vec{p}) g(\vec{k})}} \right] a^{\dag}(\vec{p}) 
a(\vec{p})+  \nonumber \\
 &  &
{1\over 4} \int d^{\nu} p \left[ - g(\vec{p}) + 
{{{\vec{p}}^2 + m^2 + {\lambda\over 2} \phi_0^2 + 
{\lambda\over 4} \int { {d^{\nu} p'}\over {(2 \pi)^3}} {1\over 
{g(\vec{p'})}}}\over {g(\vec{p})}} \right] (a^{\dag}(\vec{p}) a(-\vec{p}) +
a(\vec{p}) a(-\vec{p})) \, ,
\end{eqnarray}

\begin{equation}
\label{eq4.9}
H^{(3)}={\lambda\over {3!}} \phi_0 \int d^{\nu} x : {\eta}^3(\vec{x}) : \; ,
\end{equation}

\begin{equation}
\label{eq4.10}
H^{(4)}={\lambda\over {4!}} \phi_0 \int d^{\nu} x : {\eta}^4(\vec{x}) :
\end{equation}
where the colons mean normal ordering with respect to the vacuum $\Psi_0[\eta]$.

In the ket formalism the $n$-particle wavefunctionals are given by

\begin{equation}
\label{eq4.11}
| n \rangle = | {\vec{p}}_1 \nu_1; {\vec{p}}_2 \nu_2 ...\rangle = 
\prod_i
{ {{(2 \pi)}^{{\nu}\over 2}}  \over {V^{1\over 
2}}} {1\over {(\nu_i !)^{1\over 2}}} a^{\dag \nu_i}({\vec{p}}_i)|\,\,0 \rangle
\end{equation}
with $\sum_i \nu_i =n$. A straightforward calculation gives

\begin{equation}
\label{4.12}
\langle n|H^{(0)}| n \rangle = E[\phi_0, g(\vec{k})]
\end{equation}

\begin{equation}
\label{eq4.13}
\langle n | H^{(1)} | \rangle n =\langle n | H^{(3)} | \rangle n=0
\end{equation}

\begin{equation}
\label{eq4.14}
\langle n|\, H^{(2)}  | n \rangle = 
{1\over 2} \sum_i \nu_i\left[ g({\vec{p}}_i) + 
{ { {\vec{p}}_i + m^2 + { \lambda\over 2} \phi_0^2}\over {g({\vec{p}}_i)}} 
\right]+ {\lambda\over 8} \int{{d^{\nu} k}\over {(2 \pi)^{\nu}}} {1\over {g(
\vec{k})}} \sum_i \nu_i {1\over {g({\vec{p}}_i)}} \; .
\end{equation}
As concern $H^{(4)}$, a rather length but  elementary calculation 
shows that in the thermodynamic limit $V \rightarrow \infty$, we also have 

\begin{equation}
\label{eq4.15}
\langle n| H^{(4)} | n \rangle=0 \; .
\end{equation}
Now if we select the variational basis by minimizing $H_{00}=E_0[\phi_0,g(
\vec{k})]$, $i.e$ we impose the extremum condition Eq. (\ref{eq2.17}) 
we find that $H^{(2)}$ reduces to

\begin{equation}
\label{eq4.16}
H^{(2)}=\int d^{\nu} p \, g(\vec{p}) \, a^{\dag}(\vec{p}) \, a(\vec{p}) \; .
\end{equation}
Moreover 

\begin{equation}
\label{eq4.17}
\langle n | H | n \rangle = E[\phi_0, g(\vec{k})] + \sum_i \nu_i \, 
g({\vec{p}}_i) \; ,
\end{equation}
where $g(\vec{k})=\sqrt{{\vec{k}}^2 + \mu^2(\phi_0)}$, and $\mu(\phi_0)$
satisfies the gap equation Eq. (\ref{eq2.19}).

Equation (\ref{eq4.16}) tells us that $H^{(2)}$ is the normal ordered 
Hamiltonian of a free scalar field with mass $\mu(\phi_0)$. Moreover it is now 
clear that the off-diagonal elements of the full Hamiltonian are due to 
$H^{(1)}$, $H^{(3)}$ and  $H^{(4)}$.
As a consequence we can write (using again the gap equation):

\begin{equation}
\label{eq4.18}
H_0=H^{(0)} + H^{(2)}= E[\phi_0,g(\vec{k})]+H^{(2)},
\end{equation}

\begin{equation}
\label{eq4.19}
H_I=\int d^{\nu}x \left[ \left(\mu^2(\phi_0) \, 
\phi_0 - {{\lambda}\over 3} \phi^3_0 \right)
:\eta(\vec {x}): + {\lambda\over {3!}} \phi_0 : \eta^3(\vec {x}): + 
{\lambda\over {4!}} :\eta^4(\vec {x}): \right].
\end{equation}
We stress once again that our perturbation is given by the 
off-diagonal elements 
of the full Hamiltonian $H$ with respect to the variational basis 
$\{\Psi_n[\eta]\}$. This means that the perturbation expansion that we will 
discuss in the next Section, is not a weak coupling expansion. In other words,
 our variational procedure, 
which selects the Fock basis $\{\Psi_n[\eta]\}$, minimizes the off-diagonal
elements $H_{nm}$; so that even though the quartic selfcoupling is strong 
the perturbative expansion gives sensible results.
Finally, it is worth mentioning that the simple results Eq.(\ref{eq4.19}) for the 
perturbation Hamiltonian relies on Eq. (\ref{eq4.15}) which is valid only for 
quantum systems with an infinite number of degree of freedom.

\section{THE GENERALIZED GAUSSIAN EFFECTIVE POTENTIAL}

In the previous Section we was able to split the Hamiltonian $H$ into two 
pieces: the free Hamiltonian $H_0$ and the interaction $H_I$. If we neglect 
$H_I$ we see that the ground state of $H_0$ is the wavefunctional 
$\Psi_0[\eta]$ and the Gaussian effective potential Eq. (\ref{eq2.22})
is the ground state energy density. In other words, $V_{GEP}(\phi_0)$ is the 
lowest order term of the vacuum energy density in the perturbation expansion 
generated by $H_I$. Thus the corrections to the Gaussian effective potential
can  be readily obtained by means of the standard perturbation expansion for 
the ground state energy. For the ground state energy we may use the well-known
Brueckner-Goldstone formula \cite{B-G}:

\begin{equation}
\label{eq5.1}
E_{GS}(\phi_0)=E_0(\phi_0,g(\vec{k})) 
+ \sum^{\infty}_{n=0} \left[ \langle 0| H_I \left({1\over 
{E_0-H_0}} H_I \right)^n |0 \rangle \right]_{conn}.
\end{equation}
For instance, up to the second order in $H_I$, and using Eqs. (\ref{eq4.2})
and (\ref{eq4.3}), we have:

\begin{equation}
\label{eq5.2}
E_{GS}(\phi_0)=E_0+\sum_{n>0} (H_{00}-H_{nn})^{-1} |\langle n|H|0 \rangle|^2 \;
.
\end{equation}
Higher order terms can be analyzed by means of the so-called Goldstone diagrams 
\cite{B-G,F-W}.

However, in order to show that $E_{GS}$ in Eq. (\ref{eq5.1}) gives correctly 
the correction to the Gaussian effective potential, we must ascertain that 
it exists a state $|\Omega \rangle$ such that:

\begin{equation}
\label{eq5.3}
 E_{GS}(\phi_0)={<\Omega\,|\,H\,|\,\Omega>\over {<\Omega\,|\,\Omega>}}\; ,
\end{equation}
with the constraint 

\begin{equation}
\label{eq5.4}
\frac {\langle \Omega| \eta(\vec{x})|\Omega\rangle} 
{\langle \Omega|\Omega \rangle} =0 \; .
\end{equation}
To do this, we use the Gell-Mann-Low theorem  on the ground state 
\cite{G-M-L}. Let us consider the Hamiltonian 

\begin{equation}
\label{eq5.5}
H_{\epsilon}=H_0+H_I e^{-\epsilon |t|}, \quad \epsilon \rightarrow 0^+.
\end{equation}
Next, we introduce the temporal evolution operator

\begin{equation}
\label{eq5.6}
U_{\epsilon}(t,t_0) = \sum_{n=0}^{\infty} {{(-i)^n}\over {n!}}
\int_{t_0}^t d t_1...\int_{t_0}^t d t_n \, e^{-\epsilon (|t_1|+...+|t_n|)}
\,\,\, T[H_I(t_1)...H_I(t_n)]
\end{equation}
where $H_I(t)$ is the perturbation Hamiltonian in the interaction 
representation. 
The Gell-Mann-Low theorem says that if the following quantity exists
to all order in perturbation theory:

\begin{equation}
\label{eq5.7}
\lim_{\epsilon \to 0^+} {{U_{\epsilon}(0,-\infty) |0\rangle}\over 
{\langle 0| U_{\epsilon}(0,-\infty)|0\rangle }} \equiv 
{{|\Omega\rangle }\over {\langle 0|\Omega\rangle }} \; ,
\end{equation}
then 

\begin{equation}
\label{eq5.8}
H {{|\,\Omega>}\over {<0\,|\,\Omega>}}= E {{|\,\Omega>}\over 
{<0\,|\,\Omega>}} \; .
\end{equation}
Note that the denominator in Eqs. (\ref{eq5.7}) and (\ref{eq5.8}) is 
crucial, for the numerator and the denominator do not separately exist
as $\epsilon \rightarrow 0^+$.

From Eq. (\ref{eq5.8}) it follows

\begin{equation}
\label{eq5.9}
E=\frac {\langle \Omega | H | \Omega \rangle} 
{\langle \Omega | \Omega \rangle} \; .
\end{equation}
Now we show that $E_{GS}=E$. Indeed, from Eqs. (\ref{eq5.8}) and (\ref{eq5.5})
we get 

\begin{equation}
\label{eq5.10}
E-E_0= \frac {\langle 0 | H_I | \Omega \rangle}
{\langle 0 | \Omega \rangle}
\end{equation}
where we have taken into account that $H_0| 0 \rangle= E_0 | 0 \rangle$.

Now a standard manipulation \cite{F-W} shows that 

\begin{eqnarray}
\label{eq5.10b}
\langle 0 | H_I| \Omega \rangle=\langle 0 | \Omega \rangle
 \sum_{n=0}^\infty 
{(-i)^n\over {n!}}
\int_{-\infty}^0&d&t_1...
\int_{-\infty}^0 dt_n \; 
e^{[-\epsilon (t_1+...+t_n)]} \;  \times \nonumber \\
 & & \langle 0|T(H_I(0) H_I(t_1)...H_I(t_n)) |0\rangle_{conn} 
\end{eqnarray}
where the subscript means that we need to take into account only the connected 
terms. In order to carry out the time integrations, we consider the $nth$-order 
contribution in Eq. (\ref{eq5.10b}). Observing that 

\begin{equation}
\label{eq5.12}
H_I(t)=e^{iH_0t} H_I e^{-i H_0 t},
\end{equation}
and that the $n!$ possible time orderings give identical contributions, we get

\begin{eqnarray} 
\label{5.13}
(E-E_0)^{(n)}&=& (-i)^n \int_{-\infty}^0 d t_1 
...\int_{-\infty}^{0} dt_n e^{-\epsilon (|t_1|+...+|t_n|)} 
\langle 0 |\,H_I e^{i H_0 t_1}  \nonumber  \\  
 & & H_I e^{-i H_0(t_1-t_2)} \, ... \, H_I e^{-i H_0 (
t_{n-1}-t_n)} H_I e^{-i H_0 t_n} \,|\,0{\rangle}_{conn} \; .
\end{eqnarray}
By changing variables to relative times:
\begin{equation}
\label{eq5.14}
x_1=t_1, \,\, x_2=t_2-t_1, \, ... \, , x_n=t_n-t_{n-1}
\end{equation}
one finally obtains

\begin{equation}
\label{5.15}
(E-E_0)^{(n)}=\langle 0\,|\, H_I {1\over {E_0-H_0+in \epsilon}}H_I...
H_I {1\over {E_0-H_0+i \epsilon}} H_I \,|\, 0{\rangle}_{conn} \; .
\end{equation}
Because of the limitation to connected contributions, the limit 
$\epsilon \rightarrow 0^+$ is harmless. Hence we get
\begin{equation}
\label{eq5.16}
E-E_0=\sum_{n=0}^{\infty}\langle 0 | H_I \left(\frac {1} {E_0-H_0} H_I \right)^n
|0 {\rangle}_{conn} \; ,
\end{equation}
which shows that indeed $E=E_{GS}$.

We can finally write down the generalization of the Gaussian effective 
potential we are looking for \cite{Cea90}:

\begin{equation}
\label{eq5.17}
V_G(\phi_0)={1\over V}{<\Omega\,|\,H\,|\,\Omega>\over 
{<\Omega\,|\,\Omega>}} \; ,
\end{equation}
with the constraint 

\begin{equation}
\label{eq5.18}
\frac {\langle \Omega | \eta(\vec{x})|\Omega\rangle} 
{\langle \Omega | \Omega \rangle} =0 \; .
\end{equation}
Note that Eq. (\ref{eq5.18}) assures that the expectation value of the scalar 
field $\phi(\vec{x})$ on the state $|\Omega \rangle$ is $\phi_0$.

Several remarks are in order. Equation (\ref{eq5.16}) shows that $E$ reduces to 
$E_0$ in the zero-$th$ order due to the normal ordering of the interaction 
Hamiltonian. Thus, in that approximation $V_G(\phi_0)$ coincides with the 
Gaussian  effective potential. 

Higher order contributions to the generalized Gaussian effective potential 
$V_G(\phi_0)$ can be evaluated by the Brueckner-Goldstone formula Eq. 
(\ref{eq5.16}). In this case one deals with an expansion in terms of the 
Goldstone diagrams \cite{F-W}. However, one can do better if one uses Eqs.
(\ref{eq5.10}) and (\ref{eq5.10b}):

\begin{equation}
\label{eq5.19}
V_G(\phi_0)={1\over V} \sum_{n=0}^{\infty} \frac {(-i)^n} {n!} \int_{-\infty}^0
d t_n ... \int_{-\infty}^0 d t_n \langle 0 | T(H_I(0)H_I(t_1) ... H_I(t_n))|
0 {\rangle}_{conn}
\end{equation}
where we have performed  the harmless limit $\epsilon \rightarrow 0^+$.
Indeed a given term in Eq. (\ref{eq5.19}) can be easily evaluated by means of the 
standard Feynman diagrams. It is evident that the free Feynman propagator 
coincides with the propagator of a scalar field with mass $\mu(\phi_0)$.

It should be stressed that it is convenient  to analyze the expansion 
(\ref{eq5.19}) by means of the Feynman diagrams. Indeed, it is well
known that 
a Feynman diagram with $k$ vertices contains $k!$ Goldstone diagrams, 
corresponding to the number of permutations of the times $t_1, ... , t_k.$

The expansion Eq. (\ref{eq5.19})  gives rise to a set of vacuum Feynman diagrams. 
Let us consider the term of order $n$ in Eq. ({\ref{eq5.19}).
In the time-ordered product there are $n+1$ interaction Hamiltonian. This means 
that, unlike the usual vacuum diagrams the factorial factor $n!$ is cancelled 
by the number of permutations of the independent integration variables. As a 
consequence, Eq. (\ref{eq5.19}) allows a diagrammatic expansion of the higher 
order corrections which is amenable to a diagrammatic resummation. 

Finally, we point out that in terms of Feynman diagrams, the constraint Eq. 
(\ref{eq5.18}) sets to zero the tadpole-like diagrams  that are due to the 
linear term in the interaction Hamiltonian.

\section{CORRECTION TO THE GAUSSIAN APPROXIMATION}

In the previous Section we have introduced the generalized Gaussian effective 
potential which allows to compute in a systematic way the corrections to the 
Gaussian approximation. Presently we focus on the second order corrections. Let 
us consider the first non trivial term in the perturbative expansion 
Eq. (\ref{eq5.19}). We have

\begin{equation}
\label{eq6.1}
V_G(\phi_0)=V_{GEP}(\phi_0) + \Delta V_G(\phi_0)  
\end{equation}
with

\begin{equation}
\label{eq6.2}
\Delta V_G(\phi_0) = \frac {-i} {V} \int_{-\infty}^0 dt \, \langle 0 | T H_I(0)
H_I(t) | 0 {\rangle}_{conn}.
\end{equation}
Taking into account Eqs. (\ref{eq5.18})  and (\ref{eq4.19}) we get:

\begin{eqnarray}
\label{eq6.3}
 \Delta V_G(\Phi_0)={-i\over V}\int_{-\infty}^0 d\,t \,  \int d^{\nu} x \, 
d^{\nu} y  & \left[\left({{\lambda \phi_0}\over {3!}}\right)^2
<0\,|\,T(:\eta^3(x)::\eta^3(y):)\,|\,0>\right. +\cr
 &+ \left. \left({{\lambda}\over {4!}}\right)^2
<0\,|\,T(:\eta^4(x)::\eta^4(y):)\,|\,0> \right]
\end{eqnarray}
where $x=(0,\vec{x})$ and $y=(t, \vec{y}).$ Therefore we have (see Fig. 1):
\begin{equation}
\label{eq6.4}
\Delta V_G(\phi_0) = {-i\over V} \int_{-\infty}^0 d t \, \int d^{\nu} 
x d^{\nu} y \left[{{\lambda^2 \phi_0^2}\over {3!}} (i\, G_F(x,y))^3+ 
{{\lambda^2}\over {4!}} (i\, G_F(x,y))^4\right]
\end{equation}
where the Feynman propagator is:

\begin{equation}
\label{eq6.5}
G_F(x,y) = \int { { d^{\nu + 1} k}\over {(2 \pi)^{\nu+1}} } 
{ { e^{-ik(x-y)}} \over {k^2-\mu^2+i\epsilon} } \; .
\end{equation}
Inserting Eq. (\ref{eq6.5}) into Eq. (\ref{eq6.4}) and performing the 
time and spatial integrations we recast Eq. (\ref{eq6.4}) into:

\begin{eqnarray}
\label{eq6.6}
\Delta V_G(\phi_0)=  &-& {{\lambda^2 \phi_0^2}\over {3!}}\int 
\prod_{i=1}^3 {{d^{\nu} k_i}\over {(2 \pi)^{\nu} 2 g({\vec{k}}_i)}} \, 
{{(2 \pi)^{\nu} \delta({\vec{k}}_1+{\vec{k}}_2+{\vec{k}}_3)}\over 
{g({\vec{k}}_1)+g({\vec{k}}_2)+g({\vec{k}}_3)}}  
 \nonumber \\
 &-&  {{\lambda^2 }\over {4!}}\int \prod_{i=1}^4 
{{d^{\nu} k_i}\over {(2 \pi)^{\nu} 2 g({\vec{k}}_i)}} \,
{{(2 \pi)^{\nu} \delta({\vec{k}}_1+{\vec{k}}_2+{\vec{k}}_3+{\vec{k}}_4)}\over 
{g({\vec{k}}_1)+g({\vec{k}}_2)+g({\vec{k}}_3)+g({\vec{k}}_4)}} \; .
\end{eqnarray}                                                  
Note that the lowest order contribution in the loop expansion 
is the two-loop diagram (diagram (a) in Fig. 1) which was lost in the 
Gaussian approximation. In non-abelian gauge theories this diagram is crucial 
in order to maintain the gauge invariance in the variational Gaussian 
approximation \cite{Cea88}.

Now we discuss the second order corrections in the case of scalar fields in 
$\nu=1,2$ spatial dimensions \cite{C-T94}.

\subsection{SCALAR FIELDS IN 1+1 DIMENSION}

To start with, we consider the Gaussian effective potential in one spatial 
dimension $\nu=1$:

\begin{equation}
\label{eq6.7}
V_{GEP}(\phi_0)= \frac {m^2} {2} \phi_0^2 + {{\lambda} \over {4!}}
+I_1(\mu) - \frac {\lambda} {8} I_0^2(\mu)
\end{equation}
where 

\begin{equation}
\label{eq6.8}
I_0(\mu)={1\over 2} \int_{-\infty}^{+\infty} \frac {dk} {2 \pi} \frac {1}
{\sqrt{g(k)}} \; ,
\end{equation}

\begin{equation}
\label{eq6.9}
I_1(\mu)={1\over 2} \int_{-\infty}^{+\infty} \frac {dk} {2 \pi} 
{\sqrt{g(k)}} \; .
\end{equation}
Introducing an ultraviolet cutoff $\Lambda$, it is not difficult to see that 
the integrals (\ref{eq6.8}) and (\ref{eq6.9}) display quadratic and logarithmic 
divergences. Subtracting the energy density of the $\phi_0=0$ vacuum $V_{GEP}(
\phi_0=0)$, one is left with a logarithmic divergence which can be cured by 
renormalizing the mass. In the Gaussian approximation the renormalized mass is 
defined as \cite{Stev84}:
\begin{equation}
\label{eq6.10}
m^2_R \equiv \left.{{\partial^2 V_{GEP}(\phi_0)}\over 
{\partial \phi_0^2}}\right|_{\phi_0=0} = m^2 + {{\lambda}\over 2} 
I_0(\mu_0)=\mu_0^2
\end{equation}
where we recall that $\mu_0=\mu(\phi_0=0).$ However we shall see that 
 once we consider the 
corrections to the Gaussian approximation the prescription Eq. (\ref{eq6.10})
needs modification.

A serious problem of the variational approximation in quantum field theories
is due to the presence of ultraviolet divergences \cite{F88}.
The variational-perturbation theory discussed in this paper offers a natural
solution to the ultraviolet divergences problem which parallels the 
perturbative renormalization theory.
As a matter of  fact, we showed that the generalized Gaussian effective 
potential $V_G(\phi_0)$ is the energy density of the vacuum $|\Omega \rangle$
with scalar condensate $\phi_0 = \frac {\langle\Omega | \phi | \Omega \rangle}
{\langle \Omega| \Omega \rangle}.$
Observing that only the energy differences are of importance, our 
renormalization prescription will be to reabsorb the ultraviolet divergences
in the field theory without scalar condensate. Moreover  for $\phi_0=0$
the Hamiltonian  Eqs. (\ref{eq4.18}) and (\ref{eq4.19}) reduces to the one of a
scalar field with 
mass $\mu_0$ and normal ordered quartic selfinteraction. In the case of one and 
two spatial dimensions that theory is super-renormalizable and we only need to 
renormalize the mass. We define the renormalized mass by means of the 
zero-momentum 2-point proper vertex $\Gamma^{(2)}(p;\phi_0=0)$:
\begin{equation}
\label{eq6.11}
m^2_R=-\Gamma^{(2)}(0;\phi_0=0) \; .
\end{equation}
In the Gaussian approximation the $\phi_0=0$ Hamiltonian coincides with the 
free Hamiltonian  
of a scalar field with mass $\mu_0$.
So that Eq. ({6.11}) gives 
\begin{equation}
\label{eq6.12}
m^2_R \; = \; \mu_0^2 
\end{equation}
which agrees with Eq. (\ref{eq6.10}). In one spatial dimension Eq. 
(\ref{eq6.12}) eliminates completely the ultraviolet divergences, 
for the higher order corrections are finite. 

From Equation (\ref{eq6.12}) and the gap equation we get 
\begin{equation}
\label{eq6.13}
m^2=m^2_R - \frac {\lambda_0} {2} I_0(m_R) \; .
\end{equation}
Inserting Eq. (\ref{eq6.13}) into Eq. (\ref{eq6.7}) and using Eqs. 
(\ref{eq2.24}-\ref{eq2.27}) and (\ref{eq2.22}) one obtains \cite{Stev84}:

\begin{equation}
\label{eq6.14}
V^{1+1}_{GEP}(\phi_0)=-2 \hat {\lambda} \Phi_0^4 + 
{{x-1}\over {24 \hat {\lambda}}}\left[1+{{3 \hat {\lambda}}\over \pi} 
+ {{x-1}\over 2}
\right] 
\end{equation}
with the gap equation 

\begin{equation}
\label{eq6.15}
x-1+ { 3\over {\pi}} \ln x = 12 \hat {\lambda} \, \Phi_0^2 \; .
\end{equation}
These results have been obtained for the first time by S.J. Chang \cite{CH75}.
In Figure 2 display $V_{GEP}^{1+1}(\Phi_0)$ as a function of $\Phi_0$ for 
various values of the dimensionless coupling $\hat{\lambda}$. Several features 
are worth mentioning. Firstly, $\Phi_0$ is always a local minimum of $V_{GEP}(
\Phi_0)$. For $\hat{\lambda} < {\hat{\lambda}}_c$, with  ${\hat{\lambda}}_c
\simeq 2.5527$, the $\Phi_0=0$ vacuum is the true ground state.  On the other 
hand, for $\hat{\lambda} > {\hat{\lambda}}_c$ the ground state is for 
$\Phi_0 \neq 0$. As a consequence, a first order phase transition occurs at 
${\hat{\lambda}}_c.$  However, Chang \cite{CH76} pointed out that  the 
Simon-Griffiths theorem \cite{SG} rules out the possibility 
of a first-order phase 
transition in the one-dimensional $\lambda \phi^4$ field theory. Moreover, 
Chang \cite{CH76} showed that there is no contradiction between the existence 
of a second order transition and the Simon-Griffiths theorem. 

Remarkably, it turns out that the two loop correction, diagram (a) in Fig. 1, 
gives rise to a second order phase transition. 
Indeed that correction is finite:

\begin{equation}
\label{eq6.16}
\Delta V_G(\Phi_0)= -a \; \frac {{\hat{\lambda}}^2} {x} \; \Phi_0^2 \; \mu_0^2
\end{equation}
with

\begin{eqnarray}
\label{eq6.17}
a={3\over {\pi^2}}\int_{-\infty}^{+\infty} dx \,& dy &  
{1\over {\sqrt{(x^2+1)(y^2+1)[(x+y)^2+1]}}} \; \times \nonumber  \\ 
 &  & {1\over {[\sqrt{x^2+1}+\sqrt{y^2+1} + \sqrt{(x+y)^2+1}]}} \simeq 0.7136
\; .
\end{eqnarray}
So that in this approximation we get  
\begin{equation}
\label{eq6.18}
\frac {V_G(\Phi_0)} {\mu_0^2} = V_{GEP}^{1+1} (\Phi_0) - a 
\frac {{\hat{\lambda}}^2} {x} \Phi_0^2 \; .
\end{equation}
In figure 3 we display Eq. (\ref{eq6.18}). We see that now there is a 
second order phase 
transition at ${\hat{\lambda}}_c = \frac {1} {\sqrt{2a}} \simeq 0.8371$.
This is confirmed  by considering the mass-gap of the $\Phi_0=0$ vacuum:

\begin{equation}
\label{eq6.19}
m^2_{phys}=\mu_0^2+\Sigma(0)
\end{equation}
where $\Sigma(p)$ is the proper self-mass of the $\phi_0=0$ theory.
In the second order
approximation $\Sigma(p)$ is given by the so-called setting sun diagram 
\cite{RAMOND}. It is easy to show that 

\begin{equation}
\label{eq6.20}
\Sigma(0)= - \frac {{\lambda}^2} {3!} \int d^2 x \,  [G_E(x)]^3
\end{equation}
where $G_E(x)$ is the Euclidean Feynman propagator

\begin{equation}
\label{eq6.21}
G_E(x)= \int \frac {d^2k} {(2 \pi)^2} \frac {e^{i k x}} {k^2 + \mu_0^2} \; .
\end{equation}
A straightforward calculation gives

\begin{equation}
\label{eq6.22}
\Sigma(0)= -2 a {\hat{\lambda}}^2 \mu_0^2 \; .
\end{equation}
Thus, from Eqs. (\ref{eq6.19}) and (\ref{eq6.22}) we get 

\begin{equation}
\label{eq6.23}
\frac {m^2_{phys}} {\mu_0^2} = \left(1- \frac { {{\hat{\lambda}^2}}}
{{{\hat{\lambda}}^2_c}} \right) \; .
\end{equation}
One can easily check that in this case 

\begin{equation}
\label{eq6.23a}
m^2_{phys}=\left.
\frac {\partial^2 V_G(\phi_0)}  
{\partial \phi_0^2}   \right|_{\phi_0=0} \; .
\end{equation}
A remarkable consequence of Eq. (\ref{eq6.23a}) is that the mass 
renormalization of the Gaussian effective potential extends to $V_G(\phi_0)$
in the two-loop approximation.
Equation (\ref{eq6.23}) tells us that the $\phi_0=0$ vacuum is stable for 
$\hat{\lambda} < {\hat{\lambda}}_c$. Moreover near the critical coupling we 
have 

\begin{equation}
\label{eq6.24}
\frac {m_{phys}} {\mu_0} \sim ({\hat{\lambda}}_c - {\hat{\lambda}})^{1\over 2}
\; ,
\end{equation}
so that the correlation length $\xi={1\over {m_{phys}}}$ diverges as

\begin{equation}
\label{eq6.25}
\xi \sim ({\hat{\lambda}}_c - \hat{\lambda})^{- \nu}, \, \,\,  \, \, \, \, \, 
\nu={1\over 2} \; .
\end{equation}
Our results are in agreement with previous studies  \cite{CH76,P-R,TH}. 
However  our generalized Gaussian effective potential relies on a 
firm field-theoretical basis which allow us  to take  care of the higher order 
corrections. Moreover in our scheme there are not ambiguities in the 
renormalization of the ultraviolet divergences. 

Let us consider now the contribution due to the diagram (b) in Fig. 1. We have 

\begin{equation}
\label{eq6.26}
\Delta V_G(\phi_0)= -a \; \frac {{\hat{\lambda}}^2} {x} \; \phi_0^2 \, 
\mu_0^2 \; - \; b \; \frac 
{{\hat{\lambda}^2}} {x}
\end{equation}
where

\begin{eqnarray} 
\label{eq6.27}
b= {3\over {16 \pi^2}} \int_{-\infty}^{+\infty} dx dy &dz&
\frac {1} {\sqrt{(x^2+1)(y^2+1)(z^2+1) [(x+y+z)^2+1]}} \; \times \nonumber  \\
 & & \frac {1}  {{ {\sqrt{x^2+1}} {\sqrt{y^2+1}} {\sqrt{z^2+1}}
{\sqrt{(x+y+z)^2+1}} }} \simeq 0.0509 \; .
\end{eqnarray}
We would like to stress that now Eq. (\ref{eq6.23a}) is no longer valid. In the 
present case this does not matter due to the fact that the second order 
corrections are ultraviolet finite. However, in the case of two spatial 
dimensions these corrections are divergent. Thus, adopting the renormalization 
prescription Eq. (\ref{eq6.23a}) instead of Eq. (\ref{eq6.11})
it may lead to an incongruous result.

In Figure 4 we contrast the generalized Gaussian effective potential in the 
two-loop approximation (full lines) and in the full second order approximation 
(dashed lines). 

Few comments are in order. The order of the transition 
is not modified  by the second order three-loop correction. Moreover the 
critical coupling ${\hat{\lambda}}_c  \simeq 1.1486 $ is quite close to 
our previous value. As a matter of fact, in the critical region 
$ \hat{\lambda} \simeq 1$ the effects of the three-loop correction do not 
substantially change the shape of the potential. Therefore we can safely 
conclude that the most important contributions in the critical region are
given  by the two loop term. This suggests that higher order corrections 
do not modify the order of the transition. 

\subsection{SCALAR FIELDS IN 2+1 DIMENSIONS}

In the case of two spatial dimensions the Gaussian effective potential is 

\begin{equation} 
\label{eq6.28}
V_{GEP}(\phi_0)={{m^2}\over 2} \phi_0^2 + \frac {\lambda} {4!} \phi_0^4 + 
I_1(\mu)- \frac {\lambda} {8} I_0^2(\mu)\; ,
\end{equation}
where, now,
\begin{equation}
\label{eq6.29}
I_0(\mu)={1\over 2} \int \frac {d^2 k} {(2 \pi)^2} 
\frac {1} {\sqrt{g(\vec{k}})}
\end{equation}
\begin{equation}
\label{eq6.30}
I_1(\mu)={1\over 2} \int \frac {d^2 k} {(2 \pi)^2}  {\sqrt{g(\vec{k}})} \; .
\end{equation}
Subtracting the energy density of the $\phi_0=0$ vacuum we get 

\begin{equation}
\label{eq6.31}
V_{GEP}(\phi_0)-V_{GEP}(0)= {{m^2}\over 2} \phi_0^2 + \frac {\lambda} {4!}
\phi_0^4 + I_1(\mu)-I_1(\mu_0) - \frac {\lambda} {8} 
[I_0^2(\mu)-I_0^2(\mu_0)] \; ,
\end{equation}
which is still affected by ultraviolet divergences. 

Introducing the renormalized mass

\begin{equation}
\label{eq6.32}
m^2_R= \left. \frac {\partial^2 V_{GEP}(\phi_0)} { \partial \phi_0^2} 
\right|_{\phi_0=0} = m^2 + {{\lambda}\over 2} I_0(\mu_0) = \mu_0^2
\end{equation}
and using the gap equation

\begin{equation}
\label{eq6.33}
\mu^2(\phi_0)= m^2+ {{\lambda}\over 2} \phi_0^2 + 
{{\lambda}\over 2} I_0^2(\mu)
\end{equation}
we get the finite  result \cite{Stev84}

\begin{equation}
\label{eq6.34}
V_{GEP}(\Phi_0)= {{\Phi_0^2}\over 2} + \hat{\lambda} \Phi_0^4 - 
\frac {(\sqrt{x}-1)^2} {24 \pi} \left[ 1+ {9\over {2 \pi}} \hat{\lambda} + 2 
\sqrt{x} \right].
\end{equation}
Similarly we can rewrite the gap equation Eq. (\ref{eq6.33}) as 

\begin{equation}
\label{eq6.35}
x=1+12 \hat{\lambda} \Phi_0^2 - {3\over {\pi}} \hat{\lambda} (\sqrt{x} -1)
\end{equation}
whose explicit positive solution is:

\begin{equation}
\label{eq6.36}
\sqrt{x} = - \frac {3 \hat{\lambda}} {2 \pi} + \sqrt{ \left(1+ 
\frac {3 \hat{\lambda}} { 2 \pi}\right)^2 + 12 \hat{\lambda} \Phi_0^2} \; .
\end{equation}
In Figure 5 we show $V_{GEP}^{2+1}(\Phi_0)$ versus $\Phi_0$ for three different
values of $\hat{\lambda}$. Again we find a first order phase transition at the 
critical coupling  $\hat{\lambda} \simeq 3.0784$ \cite{Stev84}. 
Note that in two spatial 
dimensions  there are not rigorous results which could exclude a first order 
transition. Nevertheless it is important to investigate the effects of the 
second order corrections to the Gaussian effective potential. From Equation 
(\ref{eq6.4}) we have

\begin{equation}
\label{eq6.37}
\Delta V_G(\Phi_0)=-i \int_{-\infty}^0 dx \int d^2 z \left\{
\frac {\lambda^2 \phi_0^2} { 3!} [i G_F(z)]^2 + \frac {\lambda^2} {4!} 
[i G_F(z)]^4 \right\}
\end{equation}
where $z=(t, \vec{x})$. Now, observing that the Feynman propagator 
is an even function  
and performing  the Wick rotation, we obtain:

\begin{equation}
\label{eq6.38}
\Delta V_G(\Phi_0)= -{1\over 2} \int d^3 z_E \left\{ \frac {\lambda^2 \phi_0^2}
{3!} G_E^3(z_E) + \frac {{\lambda}^2} {4!} G_E^4(z_E) \right\}
\end{equation}
where \cite{G-R}

\begin{equation}
\label{eq6.39}
G_E(z_E)= \int \frac {d^3 k_E} {(2 \pi)^3} \frac {e^{-i k_E \cdot z_E}} 
{k_E^2 + \mu^2} = \frac {\mu} {{(2 \pi)}^{3\over 2} (\mu z)^{1\over 2}}
K_{1\over 2}(\mu z) \; ,
\end{equation}
with $z=|z_E|$ and $K_{1\over 2}$ is the modified Bessel function
of order ${1\over 2}$:

\begin{equation}
\label{eq6.40}
K_{1\over 2}(x)= \sqrt{\frac {\pi} {2x}}\, e^{-x} \; .
\end{equation}
Thus we get for the Euclidean Feynman propagator:

\begin{equation}
\label{eq6.41}
G_E(z_E)= \frac {e^{-\mu z}} {4 \pi z} \, \; .
\end{equation}
Introducing

\begin{equation}
\label{eq6.42}
I_3(\mu)=\int d^3 z_E \, G_E^3(z_E) = \int \frac {d^3 z_E} {(4 \pi)^3}
\frac {e^{-3 \mu z}} {z^3}
\end{equation}

\begin{equation}
\label{eq6.43}
I_4(\mu)=\int d^3 z_E \, G_E^4(z_E) = \int \frac {d^3 z_E} {(4 \pi)^4}
\frac {e^{-4 \mu z}} {z^4} \; ,
\end{equation}
we rewrite Eq. (\ref{eq6.38}) as 

\begin{equation}
\label{eq6.44}
\Delta V_G (\phi_0) = - \frac {\lambda^2 \phi_0^2} {12} I_3 (\mu) -
\frac {\lambda^2} {48} I_4(\mu) \; .
\end{equation}
From Eqs. (\ref{eq6.42}) and (\ref{eq6.43}) we see that $I_3$ and $I_4$ are 
divergent. We regularize the integrals by means of an
the ultraviolet cutoff $\epsilon \sim {1\over 
{\Lambda}}$ \cite{S-S}: 

\begin{equation}
\label{eq6.45}
I_3(\mu, \epsilon)= {1\over {16 \pi^2}} \int_{\epsilon}^{\infty} dz \,  
\frac {e^{-3 \mu z}} {z} = -{1\over {16 \pi^2}} 
[\ln (\mu \epsilon) + \ln 3 + \gamma] + \cal{O}(\epsilon)
\end{equation}

\begin{equation}
\label{eq6.46}
I_4(\mu, \epsilon)= {1\over {(4 \pi)^3}} \int_{\epsilon}^{\infty} dz \,
\frac {e^{-4 \mu z}} {z^2} = - {1\over {(4 \pi)^3}}
\left\{ {1\over {\epsilon}} + 4 \mu\left[ \ln (\mu \epsilon) + \ln4 + \gamma-1
\right] \right\}+ \cal{O}(\epsilon)
\end{equation}
where $\gamma$ is the Euler-Mascheroni constant.

Putting it all together we obtain

\begin{eqnarray}
\label{eq6.47}
V_G(\phi_0)-V_G(0) &=& {1\over 2} m^2 \phi_0^2 + {{\lambda} \over {4!}}} 
\phi_0^4 + I_1(\mu)-I_1{\mu_0) - {{\lambda}\over 8} 
[I_0^2(\mu)-I_0^2(\mu_0)] + \nonumber \\
 & & + \frac {\lambda^2 \phi_0^2} {192 \pi^2} 
[ \ln(\mu \epsilon)+\ln 3 + \gamma]
- \frac {\lambda^2} { 768 \pi^3} (\mu - \mu_0)[\ln 4 + \gamma -1] + 
\nonumber \\
 & & - \frac {\lambda^2} { 768 \pi^3} [ \mu \, \ln (\mu \epsilon)]-\mu_0 
\ln (\mu_0 \epsilon) ] \; .
\end{eqnarray}
Now we show that the logarithmic divergences are cured by renormalizing 
the mass. To this end, we observe that \cite{Stev84}

\begin{equation}
\label{eq6.48}
I_1(\mu)-I_1(\mu_0)={1\over 2}(\mu^2-\mu_0^2)I_0(\mu_0) - \frac {\mu_0^3}
{8 \pi} \left[ {1\over 3} \left( \sqrt{\frac {\mu^2} {\mu_0^2}} -1\right)^2
\left( 2 \sqrt{\frac {\mu^2} {\mu_0^2} -1} \right) \right],
\end{equation}

\begin{equation}
\label{eq6.49}
I_0(\mu)-I_0(\mu_0) = - \frac {\mu_0} {4 \pi} \left( \sqrt{\frac {\mu^2} 
{\mu_0^2}}\right).
\end{equation}
Inserting Eqs. (\ref{eq6.48}) and (\ref{eq6.49}) into Eq. (\ref{eq6.47}),
and using the gap equation Eq.(\ref{eq6.33}), we rewrite  Eq. (\ref{eq6.47})
as:

\begin{eqnarray}
\label{eq6.50}
V_G(\phi_0)-V_G(0) &=&
 {1\over 2} \mu_0^2 \phi_0^2 + \frac {\lambda} {4!} \phi_0^4 - 
\frac {\mu_0^3} {24 \pi} \left( \sqrt{\frac {\mu^2} {\mu_0^2}}-1 \right)^2
\left(2 \sqrt{\frac {\mu^2} {\mu_0^2} -1} \right) + \nonumber \\
 & & - \frac {\lambda} { 128 \pi^2} \mu_0^2 
\left(\sqrt {\frac {\mu^2} {\mu_0^2}}-1 \right)^2 + 
 \frac {{\lambda}^2 \phi_0^2} {192 \pi^2} [ \ln (\mu \epsilon) + \ln 3 + 
\gamma] \nonumber \\
 & & - \frac {\lambda^2} {768 \pi^3} [(\mu-\mu_0)(\ln 4+ \gamma -1) 
+  \mu \, \ln  (\mu \epsilon) - \mu_0 \, \ln  (\mu_0 \epsilon) ] \; .
\end{eqnarray}
As we have already discussed, the renormalized mass is 

\begin{equation}
\label{eq6.51}
m^2_R=-\Gamma(0, \phi_0=0)=\mu_0^2 +\Sigma(0) \; .
\end{equation}
In the lowest order Gaussian approximation we have $\Sigma(0)=0$
and Eq. (\ref{eq6.51}) reduces to Eq. (\ref{eq6.32}). In the second order 
approximation we must introduce a mass counterterm so that 

\begin{equation}
\label{eq6.52}
\Sigma (0) = \delta m^2 - \frac {\lambda^2} {3!} \int d^3 x_E \,  G_E^3(x_E)
\end{equation}
where the second order term is due to the setting-sun diagram. 
Explicitly, by using the previous regularization, we find

\begin{equation}
\label{eq6.53}
\Sigma(0) = \delta m^2 + \frac {\lambda^2} {96 \pi^2} [ \ln (\mu_0 \epsilon)
+\ln  3 + \gamma] + {\cal O}(\epsilon) \; .
\end{equation}
We fix the mass counterterm  by imposing that 

\begin{equation}
\label{eq6.54}
m^2_R = \mu_0^2 \; .
\end{equation}
This results in:

\begin{equation}
\label{eq6.55}
\delta m^2 = - \frac {\lambda^2} { 96 \pi^2} [ \ln (\mu_0 \epsilon) + 
\ln 3 + \gamma] \; .
\end{equation}
As a consequence, we must introduce the following  counterterm Hamiltonian
in the $\phi_0=0$ theory:

\begin{equation}
\label{eq6.56}
\delta  H = {1\over 2} \; \delta m^2 \int d^2 x \, \phi^2(\vec{x}) \; .
\end{equation}
After writing $\phi(\vec{x})= \phi_0 + \eta(\vec{x})$ and using the constraint 
Eq.(\ref{eq5.18}), it turns out that 
$\delta H$ adds to the second order generalized Gaussian 
effective potential the further contributions depicted in Fig. 6:

\begin{equation}
\label{eq6.57}
\delta V_G^{(a)} = {1\over 2} \; \phi_0^2 \; \delta m^2
\end{equation}

\begin{equation}
\label{eq6.58}
\delta V_G^{(b)} = {1\over 2} \; \delta m^2 \int \frac {d^2 k} { (2 \pi)^2} 
{1\over {2 g(\vec{k})}} \; .
\end{equation}
Now, observing that 

\begin{equation}
\label{eq6.59}
{1\over {2 g(\vec{k})}} = \int_{-\infty}^{+\infty} \frac {d k_0} {2 \pi}
\frac {1} {k_0^2+ g^2(\vec{k})} \; ,
\end{equation}
we have $(\epsilon \rightarrow 0)$:

\begin{equation}
\label{eq6.60}
\delta V_G^{(b)}= {1\over 2} \delta m^2 G_E(\epsilon) = {1\over {8 \pi}}
\delta m^2 \left[{1\over {4 \pi \epsilon}} - \mu\right] + {\cal O}(\epsilon) \;
\; .
\end{equation}
Finally, using Eq. (\ref{eq6.55}) we get:

\begin{equation}
\label{eq6.61}
\delta V_G^{(a)} = - \frac {{\lambda}^2 \phi_0^2} {192 \pi^2} 
[\ln  (\mu_0 \epsilon) + \ln 3 + \gamma]
\end{equation}

\begin{equation}
\label{eq6.62}
\delta V_G^{(b)} = - \frac {\lambda^2} { 768 \pi^3} [\ln (\mu_0 \epsilon) + 
\ln 3 + \gamma] \left[ {1\over {4 \pi \epsilon}} - \mu \right] \; .
\end{equation}
Now, it is easy to see that $\delta V_G^{(a)}$ eliminates the ultraviolet 
divergences of the two-loop second order correction, whereas $\delta V_G^{(b)}$
makes finite the three-loop second order correction. Thus we are left with the 
finite result:

\begin{eqnarray}
\label{eq6.63}
V_G(\phi_0)-V_G(0) &=&
{1\over 2} \mu_0^2 \phi_0^2 + {{\lambda}\over 4!} \phi_0^4
-\frac {\mu_0^3} { 24 \pi} \left(\sqrt{ \frac {\mu^2} {\mu_0^2}} -1 \right)^2
\left( 2 \sqrt{ \frac {\mu^2} { \mu_0^2}} - 1 \right) + \nonumber \\
 & &
-\frac {\lambda} { 128 \pi^2} \mu_0^2 \left(\sqrt{\frac {\mu^2} {\mu_0^2}-1}
\right)^2 + \frac {{\lambda}^2 \phi_0^2} { 192 \pi^2} \ln \left( \frac {\mu} { 
\mu_0}\right) + \frac {\lambda^2} { 768 \pi^3} \mu 
[ 1- \ln (\frac {\mu} { \mu_0}
) - \ln  {4\over 3}] \nonumber \\
 & &
 - \frac {\lambda^2} { 768 \pi^3} \mu_0 [ 1- \ln  {4\over 3}] \; .
\end{eqnarray}
In terms of the scaled variables Eqs. (\ref{eq2.23}) and (\ref{eq2.24})
we rewrite Eq. (\ref{eq6.63}) as:

\begin{equation}
\label{eq6.64}
V_G(\Phi_0) = V_{GEP}^{2+1} (\Phi_0) + \frac {3 {\hat{\lambda}}^2} {\pi^2}
\Phi_0^2 \ln \sqrt{x} + \frac {3} {4 \pi^3}( \sqrt{x} -1)(1-\ln{4\over 3}) 
- {3\over {4 \pi^3}} {\hat{\lambda}}^2 \sqrt{x} \, \ln \sqrt{x} \; .
\end{equation}
It is worthwhile to study separately the effects of the two second order 
corrections. If we take into account the two-loop correction we get 

\begin{equation}
\label{eq6.65}
V_G^{(a)} (\Phi_0) = V_{GEP}^{2+1} (\Phi_0) + \frac {3 {\hat{\lambda}}^2} 
{\pi^2} \Phi_0^2 \ln \sqrt{x} \; .
\end{equation}
In Figure 7 we display Eq. (\ref{eq6.65}) for $\hat{\lambda}=1,3,$ and 5.
As it is evident there is no spontaneous symmetry breaking \cite{C-T94}.
Comparing Fig. 7  with Fig. 5, we see that the two-loop correction adds to the 
Gaussian effective potential a positive contribution which is important in the 
region $\Phi_0 \sim 1$ and overcomes the negative minimum displayed by 
$V_{GEP}^{2+1} (\Phi_0)$ in that region. On the other hand considering the 
three-loop second order correction, we have

\begin{equation}
\label{eq6.66}
V_G^{(b)} (\Phi_0) = V_{GEP}^{2+1}(\Phi_0) + \frac {3 {\hat{\lambda}}^2}
{4 \pi^3} (\sqrt{x}-1) ( 1- \ln{4\over 3}) - {3\over {4 \pi^3}} 
{\hat{\lambda}}^2 \sqrt{x} \ln \sqrt{x} \; .
\end{equation}
From Fig. 8, where display Eq. (\ref{eq6.66}) for three different values of 
$\hat{\lambda}$, we deduce that the most important effects of the three-loop
second order correction is near the origin where one gets:

\begin{equation}
\label{eq6.67}
\Delta V_G^{(b)} (\Phi_0) \stackrel {\Phi_0 \rightarrow 0} {\sim} 
- {{18}\over {4 \pi^3}} \ln {4\over 3} 
\frac {{\hat{\lambda}}^3} {1+ \frac {3 \hat{\lambda}} {2 \pi}} \Phi_0^2 \; .
\end{equation}
Indeed, Fig. 8 shows that $V_G^{(b)} (\Phi_0)$ undergoes a continuous phase 
transition at ${\hat{\lambda}}_c=3.0959.$ This feature persists even for the 
full second order generalized Gaussian effective potential Eq. (\ref{eq6.64})
(see Fig. 9). Our result is in qualitative agreement with Refs. \cite{S} and
\cite{P-H-T}. However, from Fig. 9 we see that the condensation energy is very 
small. Moreover the critical coupling ${\hat{\lambda}}_c = 3.0959$ 
differs from that of the 
Gaussian effective potential by less than one percent. We feel that the only 
sound conclusion  we can draw
is the exclusion of a first-order phase transition. Note 
that unlike of what stated in Ref. \cite{P-H-T}, the absence of a broken phase 
is not in contradiction with the analysis by S.F. Magruder \cite{M} and S. 
Chang and S.F. Magruder \cite{C-M}.

We would like to conclude this rather technical Section by stressing the most 
important achievements of our analysis. Our analysis of the ultraviolet
divergences in two spatial dimensions showed that our renormalization
 procedure works up to the second order. However, it is clear that our
renormalization can be extended to the higher orders by the usual
 renormalization procedure. Thus we feel that our results put the
generalized Gaussian effective potential on the same level as the 
effective potential.

\section{THERMAL CORRECTIONS TO THE GENERALIZED GAUSSIAN EFFECTIVE POTENTIAL}

The aim of this Section is to study the thermal corrections to the generalized 
Gaussian effective potential. For reader convenience, let us  firstly recall 
the essential points of the finite temperature effective potential 
\cite{B,W,D-J}
and the finite temperature Gaussian effective potential \cite{H-S,Rod}.

Following the classical paper by L. Dolan and R. Jackiw \cite{D-J} the finite 
temperature effective potential in the one-loop approximation is given by:

\begin{equation}
\label{eq7.1}
V_{\beta}^{1-loop}(\phi_0)={1\over {2 \beta}} \sum_n \int \frac {d^{\nu} k}
{(2 \pi)^{\nu}} \ln (E^2 + \omega_n^2)
\end{equation}
where $\omega_n = 2 \pi \beta n,$ $\beta={1\over T}$, are the Matsubara's 
frequencies \cite{MATSU}, and $E^2={\vec{k}}^2 + M^2(\phi_0)$, 
$M^2(\phi_0)=m^2 + {{\lambda}\over 2} \phi_0^2$. Performing the sum over $n$ 
one finds \cite{D-J}:

\begin{equation}
\label{eq7.2}
V_{\beta}^{1-loop} (\phi_0)=\int \frac {d^{\nu} k} {(2 \pi)^{\nu}} {E\over 2}
+ {1\over {\beta}}  \int \frac {d^{\nu} k} {(2 \pi)^{\nu}}
\ln [1-e^{-\beta  E}] \; .
\end{equation}
In the right hand of Eq. (\ref{eq7.2}) the first term is the zero-temperature 
one-loop effective potential, while the second term gives the one-loop thermal 
corrections.

As concern the Gaussian effective potential at finite temperature, we shall 
follow G.A. Hajj and P.M. Stevenson \cite{H-S}.
Let us consider a system in thermal equilibrium; this means that our system has 
minimized its free energy:

\begin{equation}
\label{eq7.3}
F=-{1\over {\beta}} \ln Z \; ,
\end{equation}
where $Z$ is the partition function:

\begin{equation}
\label{eq7.4}
Z=\text{Tr}(e^{-\beta H}) \; .
\end{equation}
In order to evaluate the thermal corrections to the Gaussian effective 
potential, the authors of Ref. \cite{H-S} split the Hamiltonian as

\begin{equation}
\label{eq7.5}
H=H_0+H_I
\end{equation}
where $H_0$ is the Hamiltonian of a scalar field with variational mass $M^2$, 
while $H_I$ comprises the remainder. The variational mass is fixed by 
minimizing the free energy Eq. (\ref{eq7.3}). To do this one uses the 
thermodynamic perturbation theory to evaluate the free energy in the lowest 
order in the perturbation Hamiltonian. Writing

\begin{equation}
\label{eq7.6}
e^{- \beta(H_0+H_I)} \simeq e^{- \beta H_0} (1- \beta H_I) \; \, ,
\end{equation}
we get 

\begin{equation}
\label{eq7.7}
Z \simeq \text{Tr} e^{- \beta H_0} [1- \beta \langle H_I \rangle^{\beta}]
\end{equation}
where $\langle O \rangle^{\beta}$ means the thermal average with respect to 
$H_0$:

\begin{equation}
\label{eq7.8}
\langle O \rangle^{\beta} = \frac {\text{Tr}(e^{- \beta H_0} O )} 
{\text{Tr} (e^{-\beta H_0})} \; .
\end{equation}
From Eqs. (\ref{eq7.3}) and (\ref{eq7.7}) we obtain:

\begin{equation}
\label{eq7.9}
F=- {1\over \beta} 
\ln \,\text{Tr} \left(e^{-\beta H_0}\right) + <H_I>^{\beta}.
\end{equation}
Now observing that the thermal average involves a summation over the 
eigenstates of $H_0$, it is not too difficult to find \cite{H-S}:

\begin{equation}
\label{eq7.10}
V_{GEP}^{T} (\phi_0)={F\over V}=
I_1+I_1^{\beta}-{{\lambda}\over 8}(I_0+I_0^{\beta})
^2 + {{m^2}\over 2}\phi_0^2+{{\lambda}\over {4!}}\phi_0^4
\end{equation}
where 

\begin{equation}
\label{eq7.11}
I_1^{\beta}\equiv {1\over {\beta}}\int {{d^{\nu }k}\over (2 \pi)^{\nu}} \ln
\left(1-e^{-\beta g(\vec{k})}\right),
\end{equation}

\begin{equation}
\label{eq7.12}
I_0^{\beta}\equiv  \int{ {d^{\nu} k}\over {(2 \pi)^{\nu}}}  
{1\over g(\vec{k})}
{1\over { e^{\beta g(\vec{k}) } -1 } }
\end{equation}
with $g(\vec{k})= \sqrt{ {\vec{k}}^2 + M^2}$. It turns out that the mass $M$ 
satisfies the thermal gap-equation:
\begin{equation}
\label{eq7.13}
M^2=m^2+{{\lambda}\over 2}[I_0+I_0^{\beta} + \phi_0^2] \; .
\end{equation}
A remarkable consequence of Eq. (\ref{eq7.10}) is that the finite temperature 
Gaussian effective potential can be obtained from $V_{GEP}(\phi_0)$ with the 
substitution rule:

\begin{equation}
\label{eq7.14}
I_0 \rightarrow I_0+I_0^{\beta}
\end{equation}

\begin{equation}
\label{eq7.15}
I_1 \rightarrow I_1+I_1^{\beta} \; .
\end{equation}
The main drawback of the Hajj and Stevenson's approach is that the splitting of 
the Hamiltonian in Eq. (\ref{eq7.5}) is not natural, for the variational mass 
$M$, which is fixed by minimizing the free energy, depends on the approximation 
adopted in evaluating perturbatively the free energy. Moreover, the 
calculations of the thermal corrections beyond  the Gaussian approximation is 
very difficult. On the other hand, as we have already discussed in Section 4, in 
our approach the Hamiltonian is split into two pieces, the free Hamiltonian 
and the interaction, in a natural manner.

As a consequence the thermal corrections to the generalized  Gaussian 
effective potential can be evaluated easily by means of the familiar  
thermodynamic perturbation theory \cite{Feynman}. In the remainder of this 
Section we focus on the lowest order thermal corrections and  compare  with 
the one-loop thermal effective potential
corrections and the finite temperature  Gaussian effective 
potential.  The higher order thermal corrections will be discussed  in the next 
Section. 

In the lowest order in the perturbation we write \cite{C-T2}

\begin{equation}
\label{eq7.16}
F=-{1\over {\beta}} \ln \text{Tr} (e^{- \beta H_0}) + \langle H_I 
\rangle^{\beta} \; ,
\end{equation}
where, now, $H_0$ is given by Eq. (\ref{eq4.18}) and $H_I$ by Eq. 
(\ref{eq4.19}). Observing that the eigenstates of $H_0$ are the states $| n 
\rangle$, Eq. (\ref{eq4.11}), with eigenvalues $E_n$ given 
Eq. (\ref{eq4.17}), we have:

\begin{equation}
\label{eq7.17}
\langle H_I \rangle^{\beta} = \frac {\text{Tr} e^{- \beta H_0} H_I}
{\text{Tr} e^{-\beta H_0}} = 
\frac {\sum_n e^{- \beta E_n} \langle n | H_I| n \rangle}
{\sum_n e^{- \beta E_n}} \; .
\end{equation}
According to our definition Eq. (\ref{eq4.3}) we have $\langle n | 
H_I|n\rangle=0, $ so we end with

\begin{equation}
\label{eq7.18}
\langle H_I \rangle^{\beta}=0 \; .
\end{equation}
The calculation of the partition function $Z_0=\text{Tr} e^{- \beta H_0}$
is straightforward:

\begin{eqnarray}
\label{eq7.19} 
Z_0 &=& 
 \text{Tr} e^{- \beta H_0} = e^{-\beta E_0} \text{Tr} e^{ - \beta
 \sum_{ \vec{k}} g(\vec{k}) a^{\dagger}_{\vec{k}} a_{\vec{k}}} \; \; \nonumber \\
 &=& e^{- \beta E_0} \prod_{\vec{k}} \frac {1} {1- e^{- \beta g(\vec{k})}}
    = e^{-\beta E_0} e^{- V \int \frac {d^{\nu} k} {(2 \pi)^{\nu}} \ln[
     1-e^{-\beta g(\vec{k})}]} \; ,
\end{eqnarray}
where $g(\vec{k})=\sqrt{ {\vec{k}}^2 + \mu^2(\phi_0)}$,  $\mu^2(\phi_0)$
satisfies the zero temperature gap equation Eq. (\ref{eq2.19}), and 
$E_0=E[\phi_0, g(\vec{k})]$, Eq. (\ref{eq4.6}). The insertion of Eqs. 
(\ref{eq7.18}) and (\ref{eq7.19}) into Eq. (\ref{eq7.16})  leads to

\begin{equation}
\label{eq7.20}
V_G(\phi_0)={F\over V}= V_{GEP}(\phi_0) + {1\over {\beta}} \int \frac {d^{\nu} 
k} {(2 \pi)^{\nu}} \ln (1-e^{- \beta g(\vec{k})}) \; .
\end{equation}
Note that Eq. (\ref{eq7.20}) differs from the finite temperature Gaussian 
effective potential Eq. (\ref{eq7.16}). The difference resides in the different 
use of the gap equation. In our scheme the gap equation Eq.(\ref{eq2.19}) is fixed 
once and for all. In particular  it does not depend on the temperature. 
On the other hand, in the finite temperature Gaussian effective potential 
approach  the gap equation includes the thermal effects. The gap equation
fixes the basis to sum over in the thermal average, so that different gap
equations lead to inequivalent basis. In fact the discrepancy between our 
results Eq. (\ref{eq7.20}) and the finite temperature Gaussian effective potential 
comes from the thermal average of the interaction Hamiltonian. In our approach 
Eq. (\ref{eq7.18}) holds, whereas in Ref. \cite{H-S} $\langle H_I
\rangle^{\beta} \neq 0.$ Note that the possibility 
of non-equivalent basis is a peculiar 
feature of quantum systems with an infinite number of degrees of freedom.

It is worthwhile to compare the one-loop thermal correction to the effective 
potential with our finite temperature generalized Gaussian effective potential
Eq. (\ref{eq7.20}). Comparing  Eq. (\ref{eq7.2}) with Eq. (\ref{eq7.20})
we see that the former agrees with the latter if 

\begin{equation}
\label{eq7.21}
M^2(\phi_0) = m^2 + {{\lambda}\over 2} \phi_0^2 \rightarrow \mu^2(\phi_0) \; .
\end{equation}
Now $\mu^2(\phi_0)$ satisfies the gap equation Eq. (\ref{eq2.19}) which, from a 
diagrammatic point of view, is obtained by summing the infinite set of the 
superdaisy graphs in the zero temperature propagator. In other words, if in the 
thermal corrections of the effective potential we replace the tree level mass 
of the shifted theory with the mass $\mu^2(\phi_0)$ obtained by summing the 
superdaisy graphs at $T=0$ in the propagator, then we obtain again a free 
energy density. Up to now this remarkable result in thermal scalar field 
theories holds for the one-loop approximation. In the next Section we will show 
that it extends to higher order thermal corrections too.

Let us analyze the lowest order thermal corrections Eq.(\ref{eq7.20}) in the 
case $\nu=1,2$ \cite{C-T2}. In one spatial dimension  Eq. (\ref{eq7.20}) reads:

\begin{equation}
\label{eq7.22}
V_G^T(\Phi_0)= V_{GEP}^{1+1}+ {1\over {\pi {\hat{\beta}}^2}} 
\int_0^{\infty} dt \,  \ln \, [1-e^{- \sqrt{t^2+{\hat{\beta}}^2 x}} ]
\end{equation}
where $\hat{\beta} = \beta \mu_0$. In Figure 10 we show $V_G^T(\Phi_0)-V_G^T(0)
$ (in units of $\mu_0^2$) versus $\Phi_0$ for $\hat {\lambda} > 
{\hat{\lambda}}_c$. As we can see, the symmetry broken at $T=0$ gets restored 
for $\hat{T} > {\hat{T}}_c.$ Obviously the critical temperature depends on 
$\hat{\lambda}$. For $\hat{\lambda}=4$ we find ${\hat{T}}_c \simeq 1.27.$
It turns out that ${\hat{T}}_c$ can be estimated, within a few percent, by means 
of the high-temperature expansion of the integral in Eq. (\ref{eq7.22}). From 
the results of the Appendix A (see Eq. (\ref{A.16})) we find the following 
high-temperature expansion:

\begin{eqnarray}
\label{eq7.23}
V_G^T(\Phi_0)&=&
V_{GEP}^{1+1} (\Phi_0) - \frac {\pi} { 2 {\hat{\beta}}^2} +
\frac {\sqrt{x}} {2 \hat{\beta}} + {x\over {4 \pi}} \ln \left( 
\frac {\hat{\beta} \sqrt{x}} {4 \pi} \right) \nonumber \\
 & & - \frac {\zeta(3)} {64 \pi^3} x^2 {\hat{\beta}}^2 
+ \frac {\zeta(5)} {512 \pi^5} {\hat{\beta} }^4 x^3 + {\cal O}({\hat{\beta}}^6)
\; ,
\end{eqnarray}
where $\zeta (z)$ is the Riemann's zeta function. In Figure 10 we also show the 
high-temperature expansion Eq.(\ref{eq7.23}) (dashed lines); as we can see the 
high-temperature expansion is a very good approximation even near the critical 
temperature. Indeed, for $\hat{\lambda} = 4$ Eq.(\ref{eq7.23}) predicts a 
critical temperature which differs by less than one percent from the 
numerically estimated value.

The case of two spatial dimensions can be dealt with in a similar way. We have
:

\begin{equation}
\label{eq7.24}
V_G^T(\Phi_0)=V_{GEP}^{2+1}(\Phi_0) + {1\over {2 \pi {\hat{\beta}}^3}} 
\int_0^{\infty} dt \, t\,  \ln (1-e^{- \sqrt{t^2 + {\hat{\beta}}^2x}}) \; .
\end{equation}
In Figure 11 we display Eq. (\ref{eq7.24}) (we subtract the temperature 
dependent constant $V_G^T(0))$ for three different values of the temperature 
and $\hat{\lambda}=4$. Again  the thermal corrections lead to the expected 
symmetry restoration at high temperatures. As in the previous case 
we performed the 
high-temperature expansion of the integral in Eq. (\ref{eq7.24}). We find (see 
Appendix A):

\begin{equation}
\label{eq7.25}
V_G^T(\Phi_0)=V_{GEP}^{2+1} (\Phi_0) + \frac {x-1} {8 \pi \hat{\beta}} - 
{x\over {8 \pi \hat{\beta}}}\ln ({\hat{\beta}}^2x) + {1\over {8 \pi 
\hat{\beta}}}
\ln {\hat{\beta}}^2 + \frac {x^{3\over 2} - 1} {12 \pi} - \frac 
{\hat{\beta} (x^2 - 1)} {92 \pi} \; .
\end{equation}
However, we would like to stress that the expansion parameter in the above 
mentioned integral 
is ${\hat{\beta}}^2 x$. So that in the region $\Phi_0 \sim 1$ where 
$x \gg 1$ the 
high temperature  expansion Eq. (\ref{eq7.25}) breaks down. In Appendix A we 
develop an alternative expansion which is useful in the region 
${\hat{\beta}}^2 x \geq 1$.

To conclude this Section, it is worthwhile to perform a quantitative comparison
of our generalized Gaussian effective potential with the finite 
temperature Gaussian effective potential and the one-loop thermal effective 
potential. For 
definiteness we focus on the critical temperature as a function of the coupling 
constant  $\lambda$ in the case of two spatial dimensions. In this case the one 
loop thermal effective potential reads (assuming unit mass):

\begin{equation}
\label{eq7.26}
V_{\beta}^{1-loop}(\Phi_0)={1\over 2} \Phi_0^2 + \hat{\lambda}{\Phi_0}^4 +
{1\over {8 \pi}} (1+ 12 \hat{\lambda} \Phi_0^2) - {1\over {12 \pi}} (1- 12 
\hat{\lambda} \Phi_0^2)^{3\over 2} - {1\over {24 \pi}} + I_1^{\beta}(1+12 
\hat{\lambda} \Phi_0^2) \, .
\end{equation}
As concern the finite temperature Gaussian effective potential, the critical 
temperature can be extracted from Eqs.(\ref{eq7.10}), (\ref{eq7.11}) and
 (\ref{eq7.12}) with $\nu = 2 $.
In Figure 12 we compare the critical temperature as a function of 
$\hat{\lambda}$ (in units of $\hat{\lambda}_c$) for the three different 
potentials. We see that our finite
temperature generalized Gaussian effective potential leads to a critical
temperature which increases more slowly than for the other two potentials.
This is due to our choice of the variational basis which implies 
 $\langle H_I \rangle^{\beta} = 0 $ .

\section{FINITE TEMPERATURE DIAGRAMMATIC EXPANSION}

In the previous Section we evaluated the lowest order thermal corrections by 
means of the thermodynamic perturbation theory. Presently we would like to 
calculate the higher order thermal corrections. To do this the usual 
thermodynamic perturbation theory is useless. Instead we may follow the 
Matsubara's methods \cite{MATSU,Feynman}. In the Matsubara's scheme one deals 
with scalar fields which depend on the fictitious imaginary time $\tau$ varying 
in the interval $(0, \beta)$. If the Hamiltonian of the system in thermal 
equilibrium  can be written as $H=H_0+H_I$, then one can show that the 
corrections to the thermodynamic potential are given by (see Appendix B):

\begin{equation}
\label{eq8.1}
\Delta \Omega = - {1\over {\beta}} \; \ln \; \langle T_{\tau} \, \exp \left\{ - 
\int_0^{\beta} H_I(\tau) d \tau \right\} \; \rangle^{\beta}
\end{equation}
where $H_I(\tau)$ is the interaction Hamiltonian in the Matsubara's interaction 
representation. $T_{\tau}$ is the $\tau$-ordering operator and the thermal 
averages are done with respect to 
the free field partition function. Moreover, it turns 
out that only the connected diagrams contribute to $\Delta \Omega$:

\begin{equation}
\label{eq8.2}
\Delta \Omega = -{1\over {\beta}} \sum_{m=1}^{\infty} \frac {(-1)^m} {m!} 
\int_0^{\beta} d \tau_1 ... d \tau_m \langle T_{\tau} ( H_I(\tau_1) ... 
H_I(\tau_m) ) \rangle_{conn}^{\beta} \; .
\end{equation}
In our case, if we write 

\begin{equation}
\label{eq8.3}
V_G^T(\phi_0)=V_{GEP}(\phi_0)+{1\over {\beta}}\int {{d^{\nu} k}\over 
{(2 \pi)^{\nu}}} \ln \left(1-e^{-\beta g(\vec{k})}\right) + \Delta 
V_G^T(\phi_0) \; ,
\end{equation}
we readily get:

\begin{equation}
\label{eq8.4}
\Delta V_G^T(\phi_0)=-{1\over {\beta V}}\sum_{m=2}^{\infty}
{{(-1)^m}\over {m!}}
\int_0^{\beta} d \tau_1...\tau_m<T_{\tau}(H_I(\tau_1)...H_I(\tau_m)>_{conn}^{\beta} 
\; .
\end{equation}
Note that, due to Eq. (\ref{eq7.18}), the sum in Eq. (\ref{eq8.4}) starts  from 
$m=2$. The thermal average of the time-ordered  products is evaluated by means 
of the Wick's theorem for thermal fields \cite{A-G-D}. 
In this way we obtain the  
thermal corrections to the generalized Gaussian effective potential by means of 
the connected thermal vacuum diagrams. For instance, the second order thermal 
corrections are displayed in Fig. 13. The vertices can be extracted from the 
interaction 
Hamiltonian Eq. (\ref{eq4.19}). The solid lines in Fig. 13 are the thermal 
propagators of the free scalar fields with mass $\mu(\phi_0)$:

\begin{eqnarray}
\label{eq8.5}
G_{\beta}(\vec{x}-\vec{y}, \tau_1-\tau_2)=& <T_{\tau} \eta(\vec{x}, \tau_1)
\eta(\vec{y},\tau_2)>^{\beta}= 
 & {1\over {\beta}} \sum_{n=-\infty}^{+\infty} \int{{d^{\nu} k}\over {(2 \pi)
^{\nu}}} {{e^{i[\vec{k}\cdot (\vec{x}-\vec{y})-\omega_n(\tau_1-\tau_2)]}
\over {\omega_n^2+g^2(\vec{k})}}}
\end{eqnarray}
where $\omega_n= 2 \pi n \beta$. Note that the thermal propagator is 
periodic in the time variable with period $2 \pi \beta$.

A distinguishing feature of the graphical expansion of Eq. (\ref{eq8.4}) with 
respect to Eq. (\ref{eq5.19}) stems from the fact that the factor $(m!)^{-1}$ 
coming from the $mth$ order term is not completely cancelled by the number of 
different Wick contractions corresponding to a given graph. Consequently, a 
graph contributes to $\Delta V_G^T (\phi_0)$ in proportion to a combinatoric 
coefficient depending on the order of the graph. Moreover, in evaluating the 
contribution due to a given graph one should take care of the normal ordering 
prescription in the interaction Hamiltonian. In turns  out that the normal 
ordering in $H_I$ modifies the so-called anomalous diagrams, i.e.  the diagrams 
which vanish at zero temperature \cite{G-R-H}. For instance, in Fig. 13 the 
diagrams (b), (c) and (e) are anomalous. To see this, we note that the normal 
ordering in $H_I$ is ineffective when we consider a thermal contraction of
two scalar fields belonging to different vertices. Therefore, the normal 
ordering comes into play when we contract two fields which belong to the same 
vertex. In this case we get the following thermal average:

\begin{equation}
\label{eq8.6}
{{\tilde{G}}_{\beta}(0) \, = \, \langle T_{\tau} : \eta (\vec{x}, \tau) \eta( 
\vec{x}, \tau}): \rangle^{\beta}
\end{equation}
instead of $G_{\beta}(0)$. Taking into the account canonical commutation 
relations 
between the creation and annihilation operators it is straightforward to show 
that 

\begin{equation}
\label{eq8.7}
{\tilde{G}}_{\beta}(0)= G_{\beta}(0) - \int \frac {d^{\nu} k} {(2 \pi)^{\nu}} 
{1\over {2 g(\vec{k})}} \; .
\end{equation}
Now we observe that 

\begin{equation}
\label{eq8.8}
\int {{d^{\nu} k}\over {(2 \pi)^{\nu}}} {1\over {2 g(\vec{k})}} =
\lim_{\beta \to \infty} G_{\beta}(0) \; .
\end{equation}
Indeed, from Eq. (\ref{eq8.5}) it follows that:

\begin{equation}
\label{eq8.9}
G_{\beta}(0) = {1\over {\beta}} \sum_{n=-\infty}^{+\infty} \int
{{d^{\nu} k}\over {(2 \pi)^{\nu}}} {1\over {\omega^2_n + g^2(\vec{k})}} \; .
\end{equation}
By using the well known identity \cite{G-R}

\begin{equation}
\label{eq8.10}
\text{cotgh}(x)={1\over {\pi x}}+{{2x}\over {\pi}}\sum_{n=1}^{\infty}
{1\over {x^2+n^2}} \; , 
\end{equation}
we rewrite Eq. (\ref{eq8.9}) as:

\begin{equation}
\label{eq8.11}
G_{\beta}(0)=\int {{d^{\nu} k}\over {{(2 \pi)}^{\nu}}} {1\over {2 g(\vec{k})}}
\text{cotgh} \left[{{\beta g(\vec{k})}\over 2}\right] \; . 
\end{equation}
Finally, performing the limit $\beta \rightarrow \infty$  in Eq. (\ref{eq8.11})
we recover Eq. (\ref{eq8.8}). 

Using Eq. (\ref{eq8.11}) we rewrite 
Eq. (\ref{eq8.7}) as:

\begin{equation}
\label{eq8.12}
{\tilde{G}}_{\beta}(0)=\int {{d^{\nu} k}\over {(2 \pi)^{\nu}}} 
{1\over {2 g(\vec{k})}} \left[ \text{coth} 
\left({{\beta}\over 2} g(\vec{k})\right)-1\right] \; .
\end{equation} 
Note that ${\tilde{G}}_{\beta}(0)$ is free from ultraviolet divergences in any 
spatial dimensions. This means that the ultraviolet divergences due to the 
tadpole $G_{\beta}(0)$ are cured by normal ordering the Hamiltonian at 
$T=0$, in accordance with the well known result that the thermal corrections in 
quantum field theories are ultraviolet finite \cite{K-M}. 

We are, now, in the position of extending the result implied by Eq. 
(\ref{eq7.21}) to the higher order thermal corrections. To this end we observe 
that the higher order thermal corrections to the effective potential are given 
by Eq. (1.9) of ref. \cite{D-J}. Observing that in the imaginary time formalism 
the interaction lagrangian agrees with the interaction Hamiltonian and that the 
Gaussian functional integrations with periodic boundary conditions in Ref. 
\cite{D-J} correspond to the thermal Wick theorem, we obtain the desired result
. There are, however, two further points which need to be discussed. First, our 
interaction Hamiltonian is normal ordered at $T=0$. However, our previous 
discussion tells us that the normal ordering does not affect the thermal 
corrections, for ${\tilde{G}}_{\beta}(0)$ and $G_{\beta}(0)$ differ by a 
temperature independent term. Secondly, in Ref. \cite{D-J} there is not  the 
linear term in the shifted scalar field. This means that our substitution rule 
Eq. (\ref{eq7.21}) holds for the physically relevant on-shell thermal effective 
potential.

Let us, now, explicitly evaluate the second order thermal corrections in the 
case of one spatial dimension. From Eq. (\ref{eq8.4}) we have:

\begin{equation}
\label{eq8.13}
\Delta V_G^T(\phi_0) = - {1\over {2 \beta V}} \int_0^{\infty} d \tau_1 \,
d \tau_2 \,
\langle T_{\tau} H_I(\tau_1) H_I(\tau_2) \rangle^{\beta}_{conn} \; ,
\end{equation}
which gives rise to the diagrams  depicted in Fig. 13.

It is easy to see that graph (a) is temperature-independent. So it does not 
contribute  to $\Delta V_G^T (\phi_0)$ due to the stability condition $\langle 
\Omega | \eta | \Omega \rangle=0$. As concern the graph (b) we get:

\begin{eqnarray}
\label{eq8.14}
(b)= - \frac {\lambda \phi_0} {4 \beta V} \left( \mu^2 \phi_0 - 
{{\lambda}\over 3} \phi_0^3 \right)
\int_{-\infty}^{+\infty} dx \,  dy \int_0^{\beta} d \tau_1 d \tau_2  
&\langle& T_{\tau} \eta(x, \tau_1) \eta(y, \tau_2) \rangle^{\beta} \nonumber \\
&\langle& T_{\tau} : \eta(x, \tau_2) \eta(y, \tau_2): \rangle^{\beta} \; .
\end{eqnarray}
According to our previous discussion we obtain:

\begin{equation}
\label{eq8.15}
(b)= - \frac {\lambda \phi_0^2} {4} \left(1- \frac {\lambda} {3 \mu^2} 
\phi_0^2 \right) {\tilde{G}}_{\beta}(0) \; .
\end{equation}
In a similar way we find:

\begin{equation}
\label{eq8.16}
(c)=- \frac {\lambda^2 \phi_0^2} {8} {1\over {\mu^2}} {\tilde{G}}^2_{\beta}(0)
\; .
\end{equation}
For the graph (d) we have:

\begin{equation}
\label{eq8.17}
(d)= - \frac{\lambda^2 \phi_0^2}{12} \int_{- \frac {\beta} {2}}^{+\frac {\beta} {2}}
 d \tau \int_{-\infty}^{+\infty} dx \; G^3_{\beta} (x, \tau) \; .
\end{equation}
Using Eq. (\ref{eq8.5}) and the result (see Appendix B, Eq. (\ref{B.30}))

\begin{equation}
\label{eq8.18}
{1\over {\beta}}\sum_n{{e^{i \omega_n \tau}}\over {\omega_n^2+g^2(\vec{k})}}= 
{{e^{- g(\vec{k}) |\tau|}}\over {2 g(\vec{k})}}+{1\over {2 g(\vec{k})}}
[e^{g(\vec{k}) \tau} + e^{- g(\vec{k}) \tau}] {1\over {e^{\beta g(\vec{k})
}-1}} \; ,
\end{equation}
we get

\begin{equation}
\label{eq8.19}
(d)= - {{{\lambda}^2 \phi_0^2 }\over {48 (2 \pi)^2}}
 \int_0^{\beta\over 2} d \tau 
\int_{-\infty}^{+\infty}  {{d k_1 \, dk_2}
\over {g(k_1)g(k_2)g(k_3)}}
\prod_{i=1}^3 \left[e^{-g(k_i) \tau} + { {e^{g(k_i)\tau} + e^{-g(k_i) \tau}}
\over {e^{\beta g(k_i)} -1}}\right]
\end{equation}
where $\sum_{i=1}^{3} k_i =0.$ Finally, using again Eq. (\ref{eq8.18}) we find

\begin{equation}
\label{eq8.20}
(e)=- {{\lambda^2 }\over {32 (2 \pi)}} {\tilde{G}}^2_{\beta}(0)
\int_0^{{\beta}\over 2} d \tau \,\, 
\int_{-\infty}^{+\infty} {{d k}\over {g^2(k)}} 
\left[e^{-g(k) \tau} + { {e^{g(k)\tau} + e^{-g(k) \tau}}
\over {e^{\beta g(k)} -1}}\right]^2 \; ,
\end{equation}

\begin{eqnarray}
\label{eq8.21}
(f)=- {{\lambda^2 V}\over {384 (2 \pi)^3}}\int_0^{{{\beta}\over 2}}
d \tau \int_{-\infty}^{+\infty} & & 
{{d k_1 d k_2 d k_3 }\over 
{g(k_1) g(k_2) g (k_3) g(k_4)}} \; \times \nonumber \\
 &  & \prod_{i=1}^4  
\left[e^{-g(k_i) \tau} + { {e^{g(k_i)\tau} + e^{-g(k_i) \tau}}
\over {e^{\beta g(k_i)} -1}}\right]
\end{eqnarray}
with $\sum_{i=1}^{4}k_i=0$.

Some comments are in order. In Equations (8.19-21) the $\tau$-integration can 
be performed explicitly, while the remaining integrations over the momenta must 
be handled numerically. In the limit $\beta \rightarrow \infty (T \rightarrow 
0)$ the anomalous graphs (b), (c) and (e) go exponentially to zero due to the 
factor ${\tilde{G}}_{\beta} (0)$. On the other hand, the graphs (d) and (f) 
reduce to the zero temperature second order generalized Gaussian effective 
potential. Indeed, in that limit in Eqs. (8.19-21) only the factors 
$e^{-g(k_i) \tau}$ survive. Performing the elementary $\tau$-integration we 
obtain the zero temperature contributions. As a consequence the zero 
temperature limit of $V_G^T(\phi_0)$ reduces to $V_G(\phi_0).$ 

In the high temperature limit we find that the graphs (e) and (f) dominate:

\begin{equation}
\label{eq8.22}
(e) \stackrel {\beta \rightarrow 0}
{\sim} - \; \frac {\lambda^2}{256} \; {1\over 
{\beta^3 \mu^5}} \; \; ,
\end{equation}

\begin{equation}
\label{eq8.23}
(f) \stackrel {\beta \rightarrow 0} {\sim}
- \; \frac{\lambda^2}{1536} \; {1\over 
{\beta^3 \mu^5}} \; \; .
\end{equation}
Therefore, in the intermediate temperature region ${\hat{\beta}} \sim 1$
we expect that the main contribute to $V_G^T(\phi_0)$ comes from the graphs 
(d), (e) and (f). This is indeed the case as shown in Fig. 14 
where we display the contributions due to the second order graphs for 
$\hat{\beta}=1$.

In Figure 15 we display the finite temperature generalized Gaussian  effective 
potential (in units of $\mu_0^2$) for three different values of $\hat{T}$ and
$\hat{\lambda} > {\hat{\lambda}}_c$. We see that the symmetry broken  at $T=0$
gets  restored by increasing the temperature through a continuous phase 
transition.

We have also performed  the analysis of the second order thermal corrections in 
two spatial dimensions. The calculation are very similar to the previous case.
Moreover we find that the contributions to the second order thermal corrections
 behave similarly to the ones of the one-dimensional case. So we do not discuss 
any further this matter. 

\section{CONCLUSIONS}

In this paper we have developed a perturbation theory with a variational basis 
for selfinteracting scalar quantum field theories. Our aim was to evaluate in a 
systematic manner the corrections to the variational Gaussian approximation. In 
particular we introduced the generalized Gaussian effective potential which 
allowed to determine the corrections to the Gaussian effective potential. 

Our method has been illustrated in the case of a selfinteracting scalar fields. 
However we feel that there are no problems in extending our method to scalar 
fields with continuous internal symmetry. As a matter of fact, recently our 
approach has been applied to scalar fields with $O(2)$ internal symmetry 
\cite{N-G-P}.

One of the most serious problem of the variational approximation in quantum 
field theories is due to the apparency of the ultraviolet divergences. The 
variational-perturbation theory developed in the present paper offers a 
solution to the ultraviolet divergences problem which is similar to the 
well known perturbative renormalization theory. Indeed, starting  from the fact 
that the generalized Gaussian effective potential by definition is  the vacuum 
energy density in presence of scalar condensate, we showed that the divergences 
are cured by the counterterms of the underlying field theory without scalar 
condensate.

We would like to stress that, to our knowledge, there are no rigorous results 
on the problem of ultraviolet divergences in the variational approach to 
quantum field theories. For this reason we focused on scalar field theories in 
one and two spatial dimensions, where one only needs to renormalize the mass. In 
one spatial dimension we showed that the lowest order renormalization of the 
mass assures that the higher order corrections are finite. In the case of two 
spatial dimensions we find that the our mass renormalization procedure works up 
to the second order. However, it should be clear that our prescription can be 
extended to the higher orders without problems. 

In the second part of the paper we studied the thermal corrections to our 
effective potential. In particular in our method the Hamiltonian is split 
into a free piece and an interaction in a natural way. This allows us 
to directly use the well developed thermodynamic perturbation theory to 
evaluate 
the thermodynamic potential. A remarkable consequence of our analysis is that 
the thermal corrections to the generalized Gaussian effective potential agree
with the ones of the effective potential provided we use Eq. (\ref{eq7.21}).

Let us conclude by briefly discussing the more realistic case of scalar fields 
in three spatial dimensions. There is a growing evidence that quartic 
selfinteracting scalar field theories are trivial in four dimensional spacetime 
\cite{F-F-S}. However, recently M. Consoli and P. M. Stevenson proposed that 
the vacuum of the $(\lambda \phi^4)_4$ theory is not trivial \cite{Consoli}.
More precisely, within the Gaussian variational approximation they argued that 
the elementary excitations behave as free fields while the vacuum resembles a 
Bose condensate. 

Recently, this triviality and spontaneous symmetry breaking scenario found some 
evidence in the lattice approach \cite{A-A-C,C-C-C-F} . 
If this turns out to be the case, 
we expect that the symmetry broken at zero temperature gets restored  by 
increasing the temperature. Thus our approach to the calculation of the thermal 
corrections may be useful to investigate the nature of the thermal phase
transition. In particular it is important to ascertain if the phase transition 
is first order or continuous.

\appendix{} 
\section{}  

We are interested in the high-temperature expansion $\hat{\beta} \rightarrow 0$
of the following integral: 
\begin{equation}
\label{A.1} 
h(a^2)={1\over {\pi {\hat{\beta}}^2}} 
\int_0^{\infty} d t \, \, \ln \, \left[1- 
e^{- \sqrt{ t^2 +a^2}}\right] 
\end{equation}
where $a^2={\hat{\beta}}^2 x$ . Following Ref \cite{D-J}  we consider 
 
\begin{equation}
\label{A.2}
h^{\beta}(a^2)={{\partial h}\over {\partial a^2}} = 
{1\over {2 \pi {\hat{\beta}}^2}}
\int_0^{\infty} {{d t}\over {\sqrt{t^2+a^2} (e^{\sqrt{t^2+a^2}} - 1})} \; .
\end{equation}
To perform the high-temperature expansion of Eq. (\ref{A.2}) it is useful to 
deal with 

\begin{equation}
\label{A.3}
h_1(\epsilon,a^2)= \int_0^{\infty} dt \frac {t^{- \epsilon}} 
{\sqrt{t^2+a^2} (e^{\sqrt{t^2+a^2}} -1)} 
\end{equation}
with $\epsilon \rightarrow 0^+$. Using the identity \cite{G-R}:

\begin{equation}
\label{A.4}
\sum_{n=1}^{\infty} {y\over {y^2+n^2}} = -{1\over {2 y}}+ {{\pi}\over {2}}
\text{coth}( \pi y) \; ,
\end{equation}
we rewrite Eq. (\ref{A.3}) as 

\begin{equation}
\label{A.5}
h_1(\epsilon, a)=\int_0^{\infty} dt {{t^{- \epsilon}}\over {\sqrt{t^2+a^2}}}
\left[\sum_{n=-\infty}^{+\infty} { {\sqrt{t^2+a^2}}\over 
{t^2+a^2+4 \pi^2 n^2}} - {1\over 2}\right] \equiv I_{\epsilon}^{(1)}(a^2) +
I_{\epsilon}^{(2)}(a^2) \; . 
\end{equation}
Performing  the  change of variable $y={t\over {\sqrt{a^2+ 4 \pi^2 n^2}}}$
we rewrite the first term in the right hand of Eq. (\ref{A.5}) as:

\begin{equation}
\label{A.6}
I_{\epsilon}^{(1)}(a^2)  
= \sum_{n=-\infty}^{+\infty}{1\over {(a^2+4 \pi^2 n^2)^{{1+ \epsilon}\over 2}}}
\int_0^{\infty} d y {{y^{- \epsilon}}\over {y^2+1}} \; .
\end{equation}
The last integral can be performed to yield \cite{G-R}:

\begin{equation}
\label{A.7}
I^{(1)}_{\epsilon}(a^2)={{\pi}\over 2} {1\over {cos {{\pi}\over 2}\epsilon}}
\left \{ {1\over {a^{1+ \epsilon}}}+ 2 \sum_{n=1}^{\infty} 
{1\over {(2 \pi n)^{1+\epsilon}}}
+ 2\sum_{n=1}^{\infty} {1\over {(2 \pi n)^
{1+\epsilon}}} \; [ {1\over {(1+ {{a^2}\over {4 \pi^2 n^2}})
^{{1+\epsilon}\over 2}}} -1 ] \right \} \; .
\end{equation}
Using the definition of the Riemann's zeta function we get:

\begin{equation}
\label{A.8}
I_{\epsilon}^{(1)}(a)= {{\pi}\over {2 a}} + 2^{-1-\epsilon} \pi^{- \epsilon}
\zeta (1+\epsilon) + \tilde{I}(a)=
{1\over {2 \epsilon}} + {{\pi}\over {2 a}} + {1\over 2}(\gamma- \ln 2 \pi)+ 
\tilde{I}(a) \; , 
\end{equation}
where

\begin{equation}
\label{A.9}
\tilde{I}(a^2) ={1\over 2} \sum_{n=1}^{\infty} {1\over n} \left[\left(1+
{{a^2}\over {4 \pi^2 n^2}}\right)^{-{1\over 2}}-1\right].
\end{equation}
As concern the integral $I_{\epsilon}^{(2)} (a^2)$ in Eq. (\ref{A.5}) we have 
\cite{G-R}:

\begin{eqnarray}
\label{A.10}
I_{\epsilon}^{(2)}(a^2)= -{1\over 2} \int_0^{\infty} dx 
{ {x^{-\epsilon}}\over {\sqrt{x^2+a^2}}} =& - {1\over 2} a^{-\epsilon}
{1\over {2^{1-\epsilon}}} {{\Gamma(1-\epsilon) \Gamma ({{\epsilon}\over 
2}) }\over {\Gamma(1- {{\epsilon}\over 2})}}\cr
=& -{1\over {2 \epsilon}}+ {1\over 2} \ln \,a \; \, .
\end{eqnarray}
So that in the limit $\epsilon \rightarrow 0^+$ we obtain:

\begin{equation}
\label{A.11}
\lim_{\epsilon \rightarrow 0^+} 
h_1(\epsilon, a^2) = {{\pi}\over {2 a}} + {1\over 2} \ln \,a + 
{1\over 2}[\gamma - \ln 4 \pi+ \tilde{I}(a^2)] \; .
\end{equation}
Finally we perform the Taylor expansion of $\tilde{I}(a^2)$:

\begin{equation}
\label{A.12}
\tilde{I}(a^2)=- \frac {\zeta(3)} {16 \pi^2} a^2 + \frac {3 \zeta(5)}
{ 256 \pi^4} a^4 + {\cal O}(a^6) \; .
\end{equation}
Putting it all together we obtain:

\begin{equation}
\label{A.13}
h_1(a^2)=\frac {\pi} {2 a} + {1\over 2} \ln a + {1\over 2} 
( \gamma - \ln 4 \pi) - \frac {\zeta(3)} {16 \pi^2} a^2 + 
 \frac {3 \, \zeta(5)} {256 \, \pi^4} a^4 + {\cal O}(a^6) \; .
\end{equation}
Whence: 

\begin{equation}
\label{A.14}
h^{\beta}(a^2)= {1\over {2 \pi {\hat{\beta}}^2}} \left\{ \frac {\pi} { 2a}
+{1\over 2} \ln  a + {1\over 2} (\gamma - \ln 4 \pi) 
- \frac {\zeta(3)} {16 \pi^2} a^2 + 
 \frac {3 \, \zeta(5)} {256 \, \pi^4} a^4 \right\} + {\cal O}(a^6) \; .
\end{equation}
In order to recover $h(a^2)$ we integrate $h^{\beta}(a^2)$ in $a^2$ with the 
boundary condition

\begin{equation}
\label{A.15}
h(0)= {1\over {\pi {\hat{\beta}}^2}} \int_0^{\infty} dt \, \ln (
1-e^{-t} ) = \frac {\pi} {6 {\hat{\beta}}^2} \; .
\end{equation}
We get: 

\begin{eqnarray}
\label{A.16}
h(a^2) =  {1\over {2 \pi {\hat{\beta}}^2}} &&
 \left \{  \frac {\pi^2} {3}   + \pi a - \frac {a^2} {4} + 
 \frac {a^2} {2} \ln \left(\frac {a} { 4 \pi} \right) + \frac {\gamma} {2}
 a^2 \right. \nonumber \\
 &&  \;  - \frac {1} {32 \pi^2} \zeta (3) \, a^4  
  \left. + \; \frac {1} {256 \pi^4} \zeta(5) \, a^6 + {\cal O} (a^8) \right\}. 
\end{eqnarray}
Let us, now, evaluate the high-temperature expansion ($a \rightarrow 0$) of
the following integral:
\begin{equation}
\label{A.17}
J(a^2)=\int_0^{\infty} dt \,\, t \, \ln (1-e^{-\sqrt{t^2+a^2}})={1\over 
2}\int_0^{\infty} dy \, \ln (1-e^{\sqrt{y+a^2}}) \; . 
\end{equation}
To this end, we evaluate
 
\begin{equation}
\label{A.18}
J'(a^2)={{d J}\over {d a^2}}={1\over 4}\int_0^{\infty}d y 
{1\over {\sqrt{y+a^2}}}{1\over {e^{-\sqrt{y+a^2}}-1}}
\end{equation}
Using the identity Eq.~(\ref{A.4}) we write

\begin{equation}
\label{A.19}
J'(a^2)= \lim_{\epsilon \to 0^+} 
[K_{\epsilon}^1(a^2)+K_{\epsilon}^2(a^2)] 
\end{equation}
where
\begin{equation}
\label{A.20}
K_{\epsilon}^1 ={1\over 4}\sum_{-\infty}^{+\infty} \int_0^{\infty} dy 
{{y^{-\epsilon}}\over {y+a^2+4 \pi^2 n^2}} \; , 
\end{equation}
\begin{equation}
\label{A.21}
K_{\epsilon}^2(a^2)=-{1\over 8} \int_0^{\infty} dy {{y^{-\epsilon}}\over 
{\sqrt{y+a^2}}} \; . 
\end{equation}
To evaluate $K_{\epsilon}^{(1)} (a^2)$ we proceed as we did for 
$I_{\epsilon}^{(1)} (a^2).$ We obtain 
\begin{eqnarray}
\label{A.22}
K_{\epsilon}^{(1)}(a^2)&=& {1\over {4 \epsilon}}\left[{1\over {a^{2 \epsilon}}}
+ {2\over {(4 \pi^2)^{\epsilon}}} \zeta(2 \epsilon) - {{2 a^2}\over 
{(4 \pi^2)^{1+\epsilon}}} \epsilon \, \zeta(2)\right] \nonumber \\
 &=&-{1\over 2}\left[\ln \, a + {{a^2}\over {4 \pi^2}}\zeta(2) \right] + 
  {\cal O}(\epsilon) \; .
\end{eqnarray}
As concern $K_{\epsilon}^{(2)} (a^2)$ we find 
\begin{equation}
\label{A.23}
K_{\epsilon}^{(2)}(a^2)=-{{a^{1-2\epsilon}}\over 8} B(1- \epsilon, \epsilon 
-{1\over 2})= {a\over 4} + {\cal O}(\epsilon) \; . 
\end{equation}
Using $\zeta(2)=\frac {\pi^2} {6}$ we obtain
\begin{equation}
\label{A.24}
J'(a^2)=-{1\over 2}\, \ln \,a+{a\over 4} - {{a^2}\over {48}} \; . 
\end{equation}
Integrating in $a^2$ we are led to 
\begin{equation}
\label{A.25}
J(a^2)= J(0) + {1\over 4}a^2 - {1\over 4} a^2\, \ln  \, a^2 +{{a^3}\over 6}-
{{a^4}\over {96}} \; .
\end{equation}
where
\begin{equation}
\label{A.26}
J(0) = \int_0^{\infty} dt \, t \, \ln (1- e^{-t}) = - \zeta(3)
\end{equation}
It is useful to perform also the low-temperature expansion ($a \rightarrow 
\infty$) of $J(a^2)$. To do this we note that:  
\begin{equation}
\label{A.27}
J(a^2)=-{1\over 2} \sum_{n=1}^{\infty} \int_0^{\infty} dy \; \; \frac 
{e^{-n \sqrt{y+a^2}}} {n} \; .
\end{equation}
Changing the integration variable we get
\begin{equation}
\label{A.28}
J(a^2)=- \frac {a^2} {2} \sum_{n=1}^{\infty} {1\over n} \int_0^{\infty} dt \, 
e^{-n a \sqrt{t+1}} = -a \sum_{n=1}^{\infty} \frac {e^{-na}} {n^2} \; 
- \; \sum_{n=1}^{\infty} \frac {e^{-na}} {n^3} \; .
\end{equation}
This last expression can be used to approximate $J(a^2)$ for $a \geq 1$.

\section{ }  

For reader convenience we briefly discuss the thermodynamic perturbation theory
in the Matsubara's scheme \cite{F-W,A-G-D}. Let us suppose that the 
Hamiltonian of our thermodynamic system can be written as

\begin{equation}
\label{B.1}
H \, = \, H_0 \, + \, H_I \; .
\end{equation}
We are interested in evaluating the thermodynamic potentials perturbatively in 
$H_I$. To this end we introduce the ${\cal S}$-matrix:
\begin{equation}
\label{B.2}
e^{H \tau} = { \cal S}^{-1}({\tau}) e^{H_0 \tau}, \,\,\,\, 0 \leq \tau \leq 
\beta \; .
\end{equation}
Let us consider the field operators in the Matsubara's interaction representation:
\begin{equation}
\label{B.3}
\phi(\vec{x},\tau)=e^{H_0 \tau} \phi(\vec{x}) e^{-H_0 \tau} \; .
\end{equation}
In this representation the interaction Hamiltonian reads:

\begin{equation}
\label{B.4}
H_I(\tau)=e^{H_0 \tau} H_I e^{-H_0 \tau} \; , 
\end{equation}
while

\begin{equation}
\label{B.5}
H_0(\tau) = H_0 \; .
\end{equation}
The solution of Eq. (\ref{B.2}) is well known:

\begin{equation}
\label{B.6}
{ \cal S}(\tau)= T_{\tau} \; exp\left[- \int_0^{\tau} H_I (\tau') d\tau'
 \, \right ] \;.
\end{equation}
Let us evaluate the thermodynamic potential $\Omega$:
\begin{equation}
\label{B.7}
e^{-\beta \Omega} = \text{Tr} (e^{ \beta H}) \; .
\end{equation}
From Eqs. (\ref{B.2}) and (\ref{B.7}) we get 
\begin{equation}
\label{B.8}
\Omega = -{1\over {\beta}} \ln  \, \text{Tr} \left(e^{-\beta H_0} 
{\cal S}(\beta) \right) \; .
\end{equation}
Defining 

\begin{equation}
\label{B.9}
\Omega_0 = - {1\over {\beta}} \; \ln \; \text{Tr} (e^{- \beta H_0})
\end{equation}
we get
\begin{equation}
\label{B.10}
\Omega - \Omega_0 = -{1\over {\beta}} \ln \, 
{{\text{Tr} \, e^{- \beta H_0} {\cal S}(\beta)}\over 
{\text{Tr}\,{e^{-\beta H_0}}}} \; \; .
\end{equation}
Whence
\begin{equation}
\label{B.11}
\Delta \Omega = \, - \, {1\over {\beta}} \; \ln \; \langle {\cal S}(\beta)
\rangle^{\beta} \; .
\end{equation}
Using Eq. (\ref{B.6}) we rewrite (\ref{B.11}) as

\begin{equation}
\label{B.12}
\Delta \Omega = - {1\over {\beta}} \left\{
\ln \sum_{m=1}^{\infty} \frac {(-1)^m}
{m!} \int_0^{\beta} d \tau_1 ... d \tau_m \langle T_{\tau} (H_I(\tau_1) ...
H_I(\tau_m))\rangle^{\beta}\right\}.
\end{equation}
One can show that \cite{A-G-D}:

\begin{equation}
\label{B12a}
\Delta \Omega = - {1\over {\beta}} 
\sum_{m=1}^{\infty} \frac {(-1)^m} 
{m!} \int_0^{\beta} d \tau_1 ... d \tau_m \langle T_{\tau} (H_I(\tau_1) ...
H_I(\tau_m))\rangle^{\beta}_{conn} \; .
\end{equation}
This last equation has been used in Section 8.

We would like, now, to discuss the thermal propagator of a free 
scalar field with mass $m$. From the well known expansion:

\begin{equation}
\label{B.13}
\phi(\vec{x},0)={1\over {\sqrt{V}}} \sum_{\vec{p}}[e^{+ i \vec{p} \cdot 
\vec{x}} \, a_{\vec{p}} + e^{- i \vec{p} \cdot \vec{x}} a_{\vec{p}}^{\dag}]
\end{equation}
we readily obtain:

\begin{equation}
\label{B.14}
\phi(\vec{x}, \tau) = {1\over {\sqrt{V}}} \sum_{\vec{p}}
\left[a_{\vec{p}} e^{i \vec{p} \cdot \vec{x} - E_p \tau} +
a_{\vec{p}}^{\dag} e^{-i \vec{p} \cdot \vec{x} + E_p \tau}\right],
\end{equation}
where $E_p=\sqrt{{\vec{p}}^2 + m^2}$, and we used:

\begin{equation}
\label{B.15}
e^{H_0 \tau} a_{\vec{p}} \, e^{-H_0\tau} = a_{\vec{p}} \, e^{-E_p \tau}
\end{equation}

\begin{equation}
\label{B.16}
e^{H_0 \tau} a_{\vec{p}}^{\dag} \, e^{-H_0\tau} = 
a_{\vec{p}}^{\dag} \,  e^{-E_p \tau}.
\end{equation}
We are interested in the thermal propagator:

\begin{equation}
\label{B.17}
\langle \; T_{\tau} \; \phi(\vec{x},\tau) \; \phi(0) \; 
 \rangle^{\beta} = G_{\beta} (\vec{x}, \tau) \; .
\end{equation}
Using Eq. (\ref{B.14}) and 

\begin{equation}
\label{B.18}
\langle a_{{\vec{p}}_1}a^{\dag}_{{\vec{p}}_2}\rangle^{\beta}
={{ \delta_{{\vec{p}}_1 
{\vec{p}}_2}}\over {1- e^{-\beta E_{{p}_1} }}}
\end{equation}
\begin{equation}
\label{B.19}
\langle a^{\dag}_{{\vec{p}}_1}a_{{\vec{p}}_2}\rangle^{\beta}
={{ \delta_{{\vec{p}}_1 
{\vec{p}}_2}}\over { e^{\beta E_{{p}_1}} - 1 }} \; \; ,
\end{equation}
we obtain 

\begin{equation}
\label{B.20}
G_{\beta}(\vec{x}, \tau) = {1\over V} \sum_{\vec{p}}
{1\over {2 E_p}}\left[ 
{ {e^{i \vec{p} \cdot \vec{x}-E_p \tau }} \over 
{1- e^{-\beta E_p}}} +
{ {e^{-i \vec{p} \cdot \vec{x}-E_p \tau }} \over 
{ e^{\beta E_p}-1}} \right] \; .
\end{equation}
Now we observe that

\begin{equation}
\label{B.21}
{{e^{-E_p \tau}}\over {1-e^{-\beta E_p}}}= -
\int_{\Gamma} {{d \, z}\over {2 \pi i}} 
{{e^{- z \tau}}\over {(1-e^{-\beta z})(z-E_p)}}
\end{equation}
where the integral in the complex z-plane is on the contour $\Gamma$ shown in 
Fig. 16. The integrand in Eq. (\ref{B.21}) goes to zero exponential when $|z| 
\rightarrow \infty$. Thus we deform the contour $\Gamma$ in $\Gamma'$ 
(see Fig. 16). Applying the Cauchy's integral theorem we get:
\begin{equation}
\label{B.22}
\frac{e^{-E_p \tau}}{1-e^{- \beta E_p}} = {1\over {\beta}} \sum_n \frac 
{e^{- i \omega_n \tau}} { E_p - i \omega_n}
\end{equation}
where $\omega_n=  2 \pi n  \beta$.  The other term 
in Eq. (\ref{B.20}) can 
be dealt with in a similar way. Thus we obtain:

\begin{equation}
\label{B.23}
G_{\beta}(\vec{x}, \tau) = {1\over {\beta}} \sum_n {1\over V} \sum_{\vec{p}}
{1\over {2 E_p}} \left[ \frac {e^{i \vec{p} \cdot \vec{x} - i \omega_n \tau}}
{E_p - i \omega_n} + \frac {e^{- i \vec{p} \cdot \vec{x} + i \omega_n \tau}}
{E_p - i \omega_n} \right] \; .
\end{equation}
Finally, observing that ${1\over V} \sum_{\vec{p}} \rightarrow \int \frac 
{d^{\nu} p} {(2 \pi)^{\nu}}$ and performing the change  of variables
$\vec{p} \rightarrow - \vec{p}$, $\omega_n \rightarrow - \omega_n$ in the 
second term in the right hand of Eq.  (\ref{B.23}), we get:

\begin{equation}
\label{B.24}
G_{\beta}(\vec{x}, \tau)={1\over {\beta}} \sum_n \int
{{d^3 p}\over {(2 \pi)^3}} {{e^{- i \omega_n \tau +i \vec{p} \cdot \vec{x}}}
\over {E^2_p+\omega_n^2}} \; \; .
\end{equation}
In the following we need to evaluate the sum:

\begin{equation}
\label{B.25}
{1\over {\beta}} \sum_n \frac {e^{i \omega_n \tau}} {\omega^2_n + E^2_p} \; \, .  
\end{equation}
To do this, we use the Sommerfeld-Watson transform \cite{Somm}:

\begin{eqnarray}
\label{B.26}
{1\over {\beta}} \sum_{n=-\infty}^{+ \infty} f(z=i\omega_n)
&=& {1\over {2 \pi i}} 
\int_{-i \infty}^{+i \infty} d z f(z) + \nonumber \\
 & &  {1\over {2 \pi i}}\int_{-i \infty+\epsilon}^{+i \infty +\epsilon}
[f(z)+f(-z)] {1\over {e^{\beta z} -1}}
  \equiv A_1+A_2\,.
\end{eqnarray}
In our case 

\begin{equation}
\label{B.26a}
f(z)=-\frac {e^{z \tau}} {z^2 - E_p^2}\,.
\end{equation}
Let us consider, firstly, $A_1$. We have 

\begin{equation}
\label{B.27}
A_1= - {1\over {2 \pi i}} \int_{- i \infty}^{+i\infty} d z \frac
{e^{z \tau}} {z^2-E^2_p} \, \; .
\end{equation}
If $\tau > 0$ we close the integration contour in the semiplane 
$Re\,z<0,$ 
while for $\tau<0$ the contour is closed in the semiplane $Re\, z>0$. In this 
way, by applying the residue theorem we obtain:

\begin{equation}
\label{B.28}
A_1= \frac {e^{- E_p | \tau|}} {2 E_p} \, \; . 
\end{equation}
In the same way we get
\begin{equation}
\label{B.29}
A_2= {{e^{E_p \tau}+e^{-E_p \tau}}\over {2 E_p}} 
{1\over {e^{\beta E_p}-1}} \, \; . 
\end{equation}
Combining Eqs. (\ref{B.28}) and (\ref{B.29}) we obtain the desired result:
\begin{equation}
\label{B.30}
{1\over {\beta}} \sum_n {{e^{i \omega_n \tau}}\over {\omega_n^2 + E_p^2}} =
{{e^{- E_p |\tau|}}\over {2 E_p}} +
{{e^{E_p \tau}+e^{-E_p \tau}}\over {2 E_p}} {1\over {e^{\beta E_p}-1}} \, \; . 
\end{equation}
\newcommand{\InsertFigure}[2]{\newpage\begin{center}\mbox{%
\epsfig{bbllx=2.5truecm,bblly=3.8truecm,bburx=19.5truecm,bbury=28.truecm,%
height=21.truecm,figure=#1}}\end{center}\vspace*{-.50truecm}%
\parbox[t]{\hsize}{\small\baselineskip=0.5truecm\hskip0.5truecm #2}}
\newcommand{\InsertFigureBis}[2]{\newpage\begin{center}\mbox{%
\epsfig{bbllx=2.5truecm,bblly=3.8truecm,bburx=19.5truecm,bbury=28.truecm,%
height=18.truecm,figure=#1}}\end{center}\vspace*{2.50truecm}%
\parbox[t]{\hsize}{\small\baselineskip=0.5truecm\hskip0.5truecm #2}}

\InsertFigure{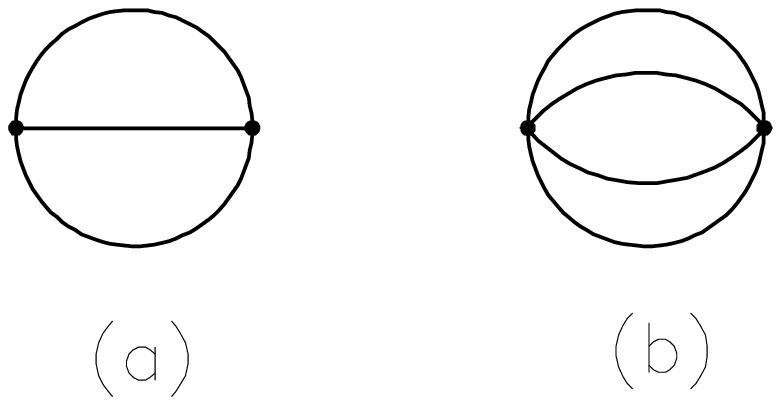}%
{FIG.~1        Second order corrections to the Gaussian effective potential.}
\InsertFigure{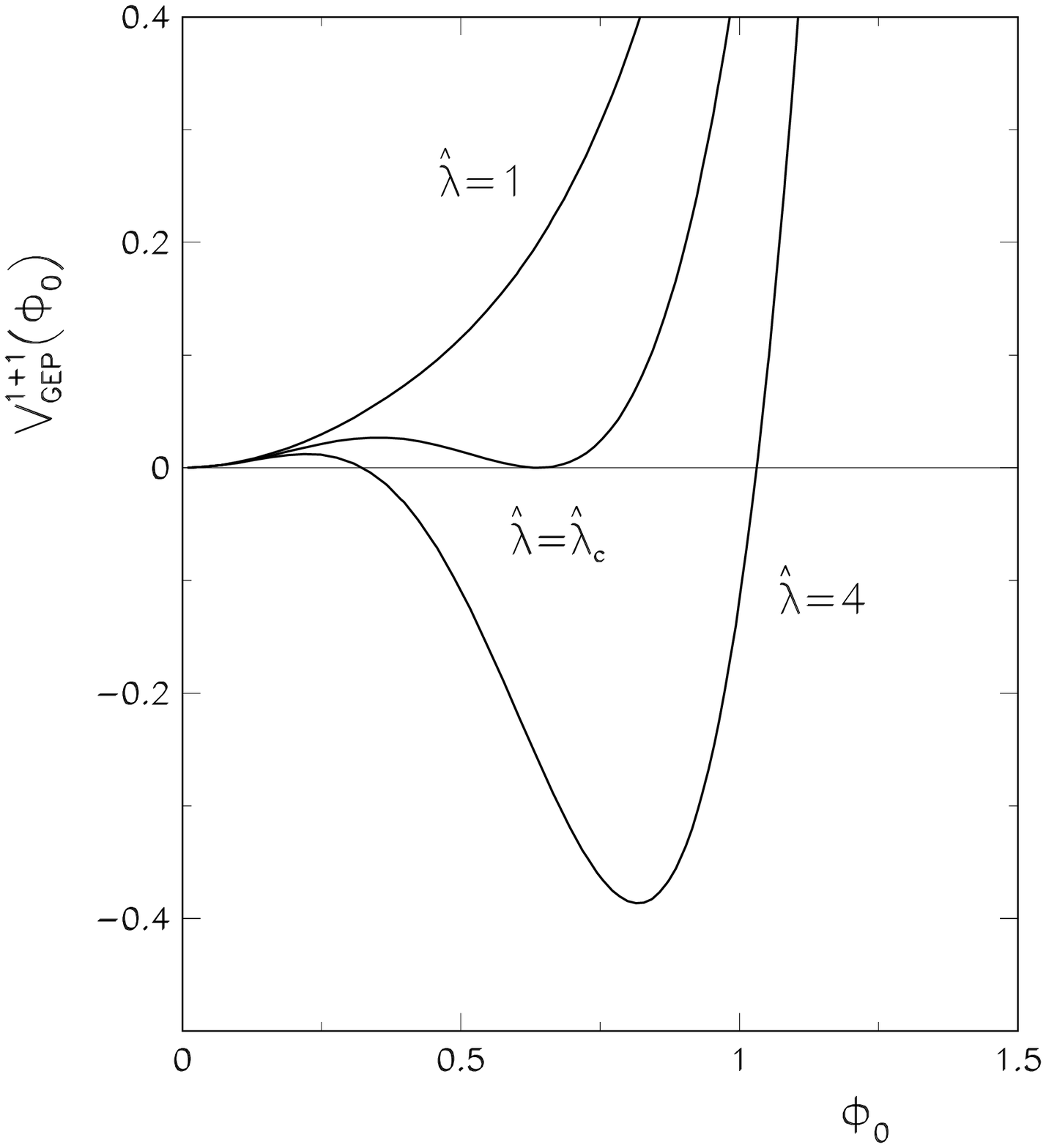}%
{FIG.~2         The Gaussian effective potential in one spatial dimension for 
                 three diferent values of $\hat{\lambda}$. } 
                                                              
\InsertFigure{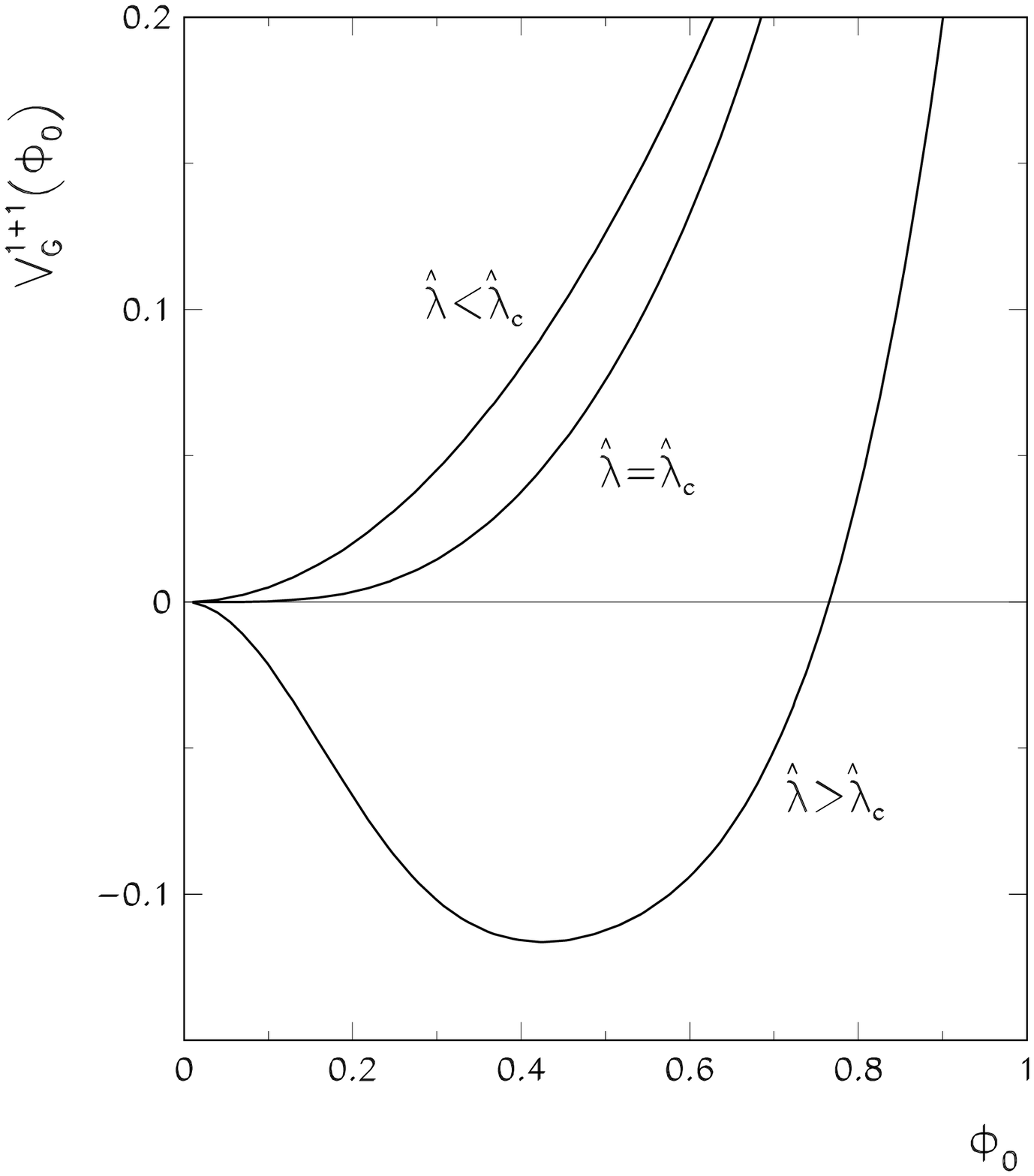}%
{FIG.~3         The two-loop Gaussian effective potential for $\nu=1$ and
                three different values of $\hat{\lambda}$.}
\InsertFigure{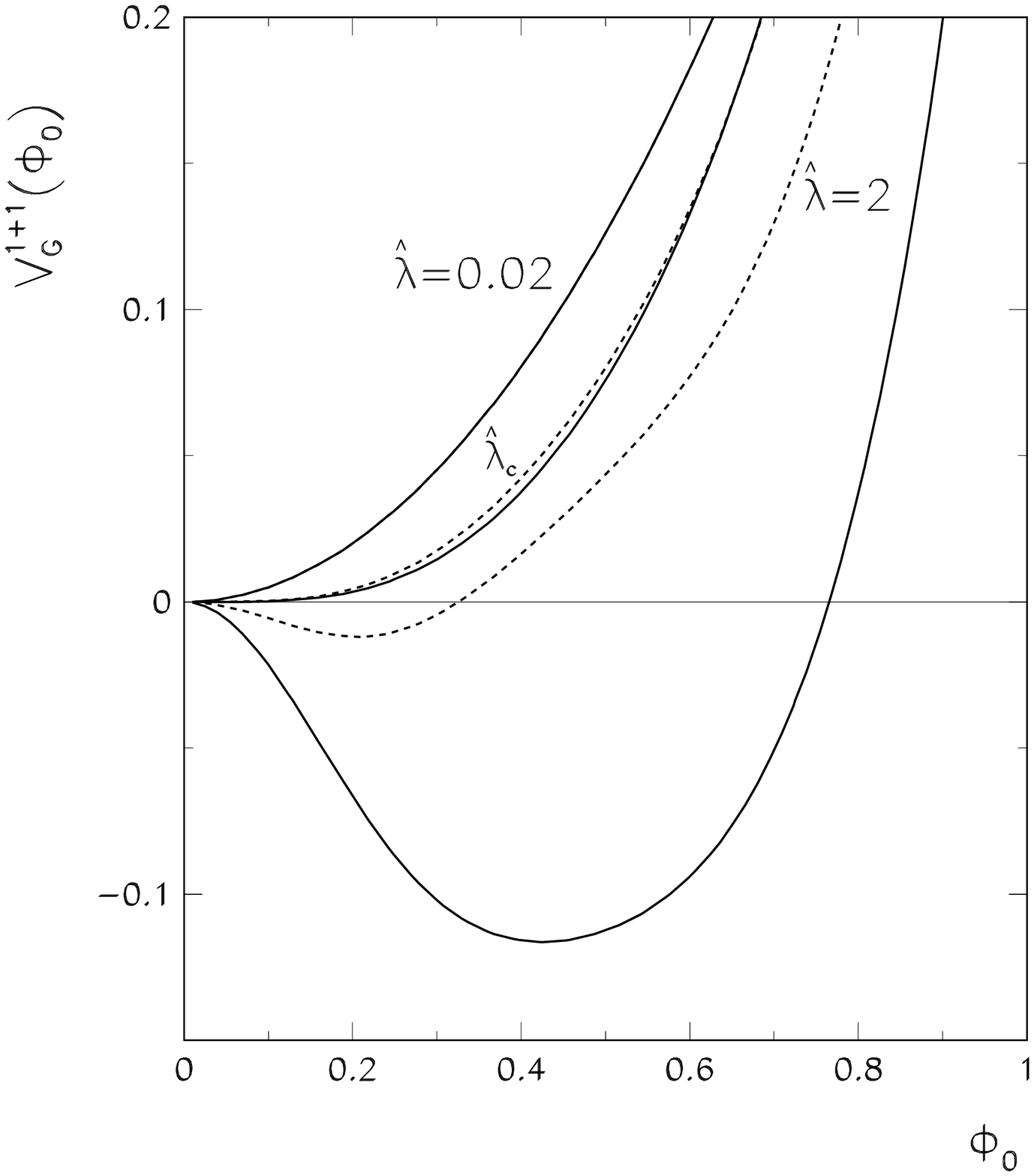}%
{FIG.~4        The two-loop (full lines) and the second order (dashed lines)
               generalized Gaussian effective potential for $\nu=1$ and three 
               different values of $\hat{\lambda}$.}
\InsertFigure{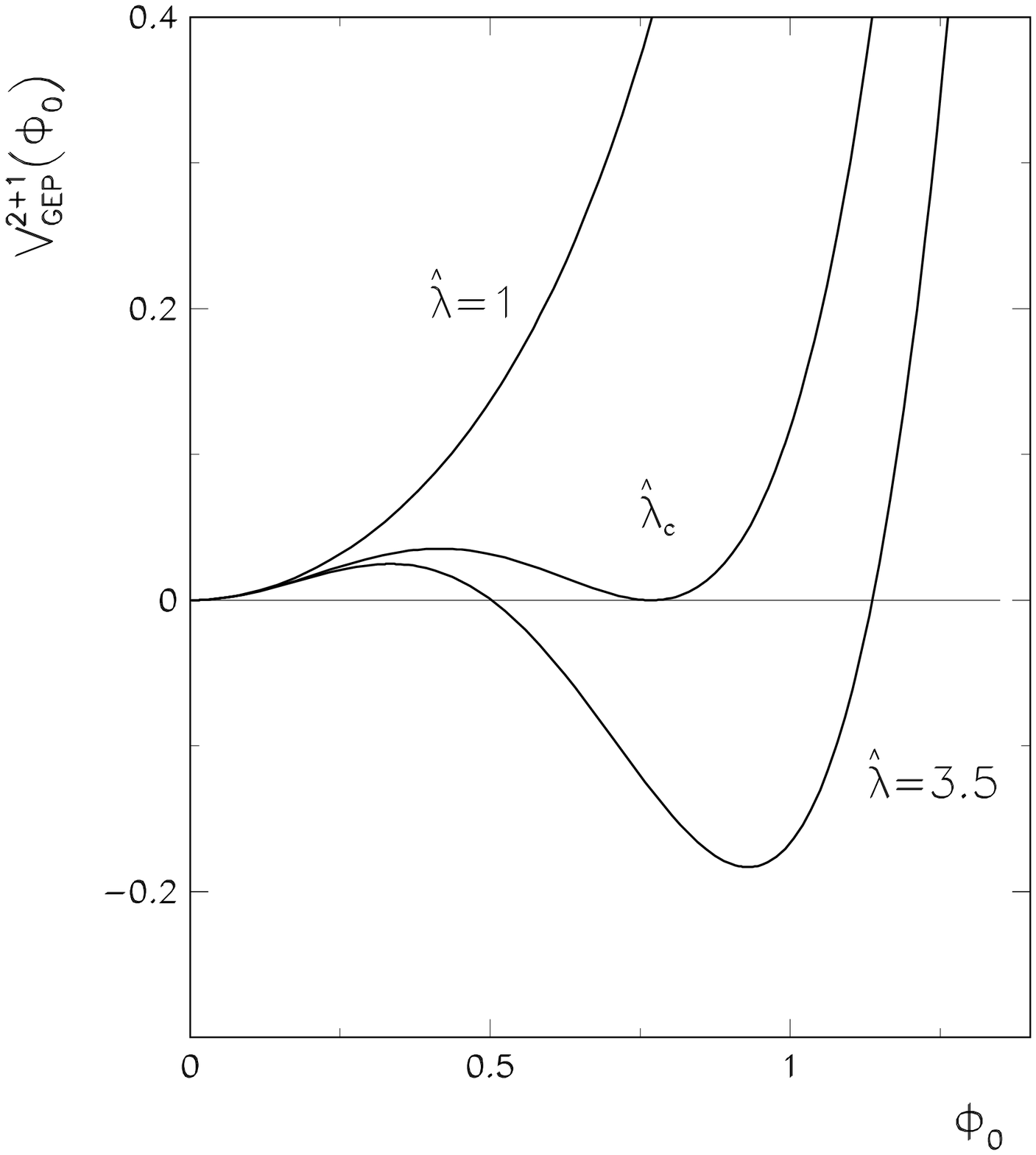}%
{FIG.~5        The Gaussian effective potential in two spatial dimensions for 
               three different values of $\hat{\lambda}$.}
\InsertFigure{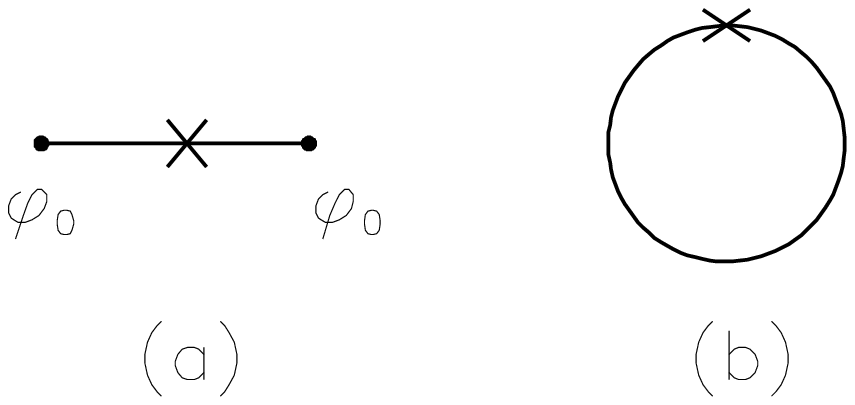}%
{FIG.~6        Mass counterterm contributions to the generalized Gaussian 
               effective potential in the second order approximation.}
\InsertFigure{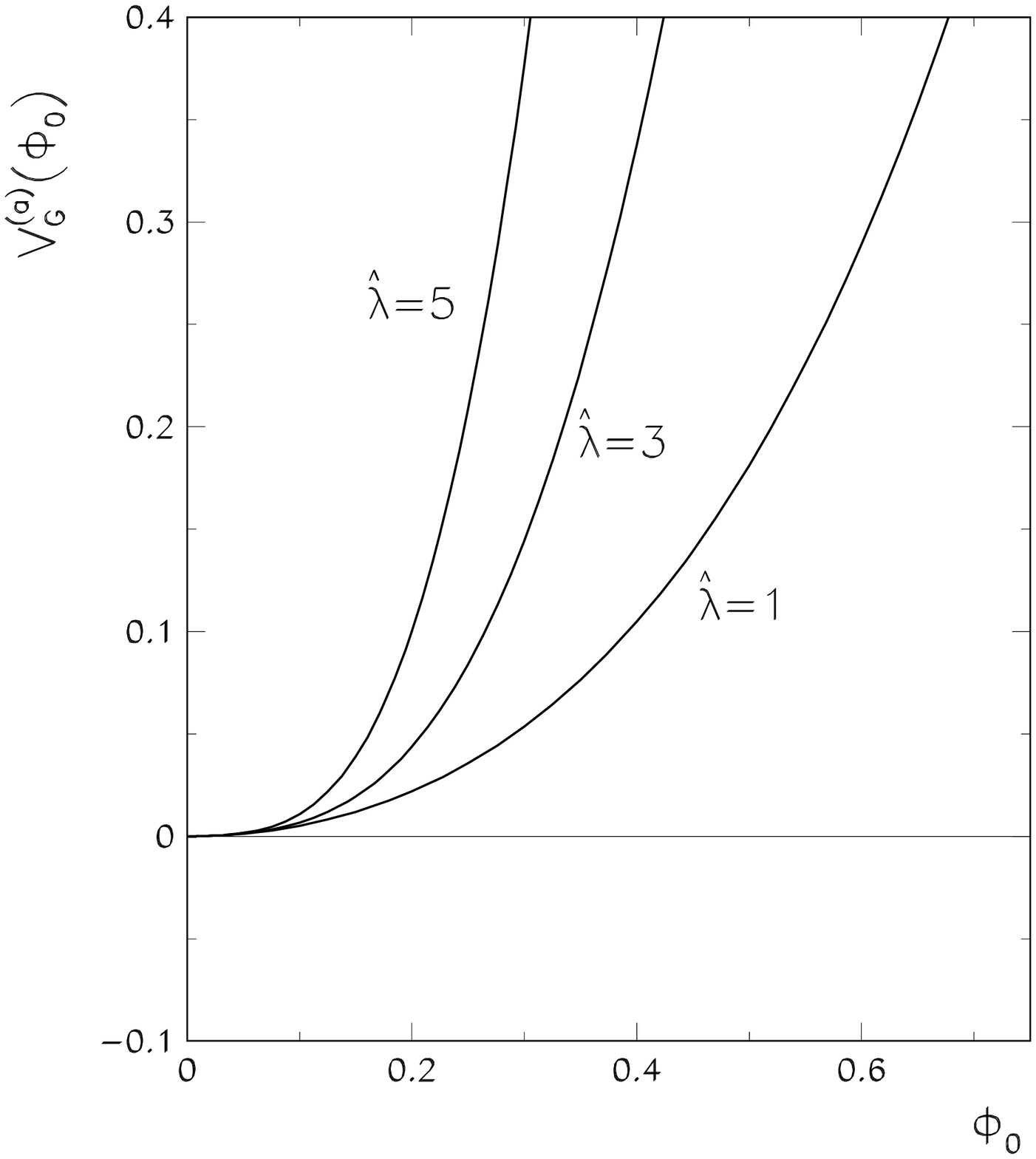}%
{FIG.~7        The generalized Gaussian effective potential for $\nu=2$ and 
               three different values of $\hat{\lambda}$ with the two-loop 
               corrections.}
\InsertFigure{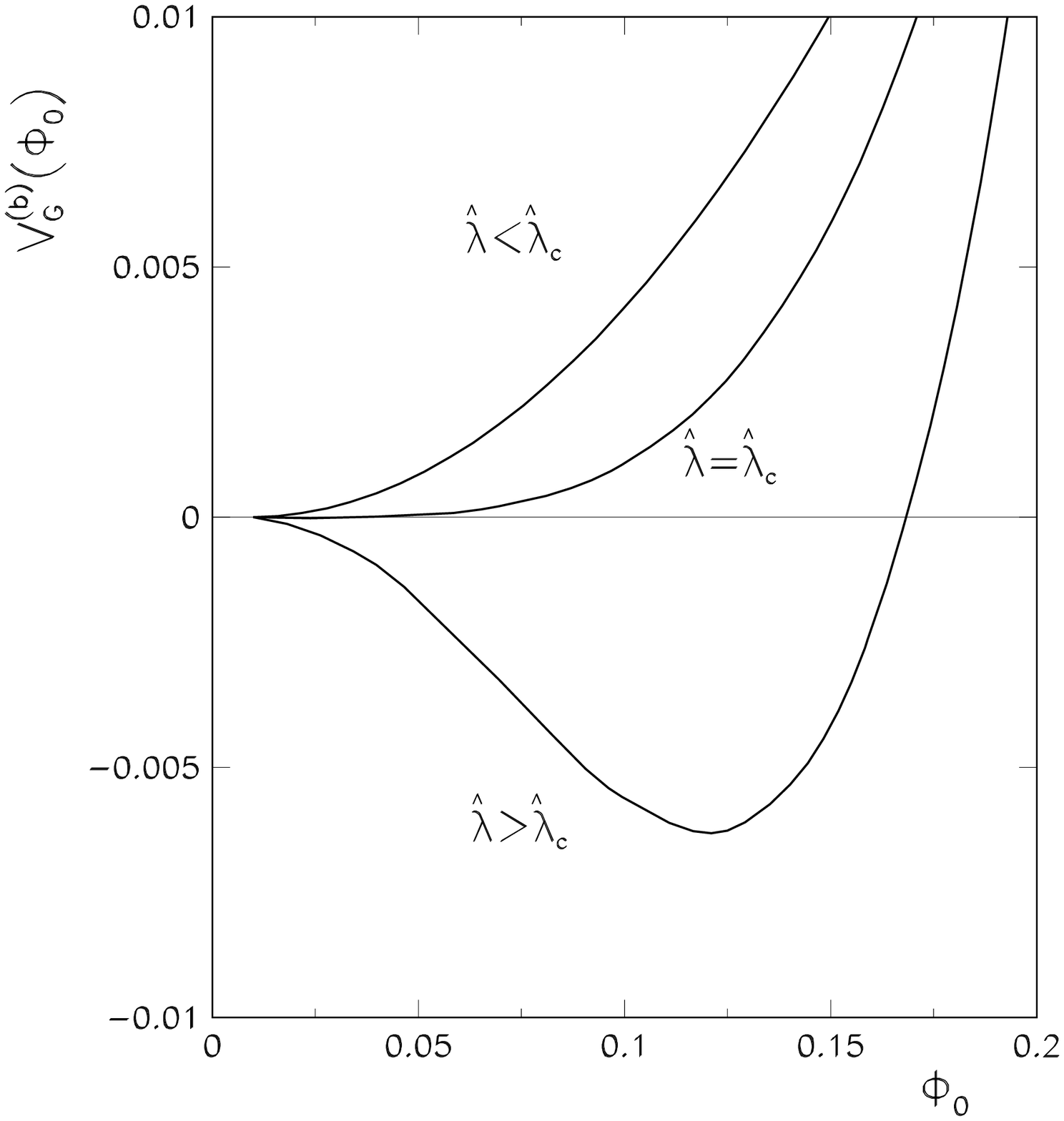}%
{FIG.~8        The generalized Gaussian effective potential for $\nu=2$ and 
               three different values of $\hat{\lambda}$ with the three-loop 
               corrections.}
\InsertFigure{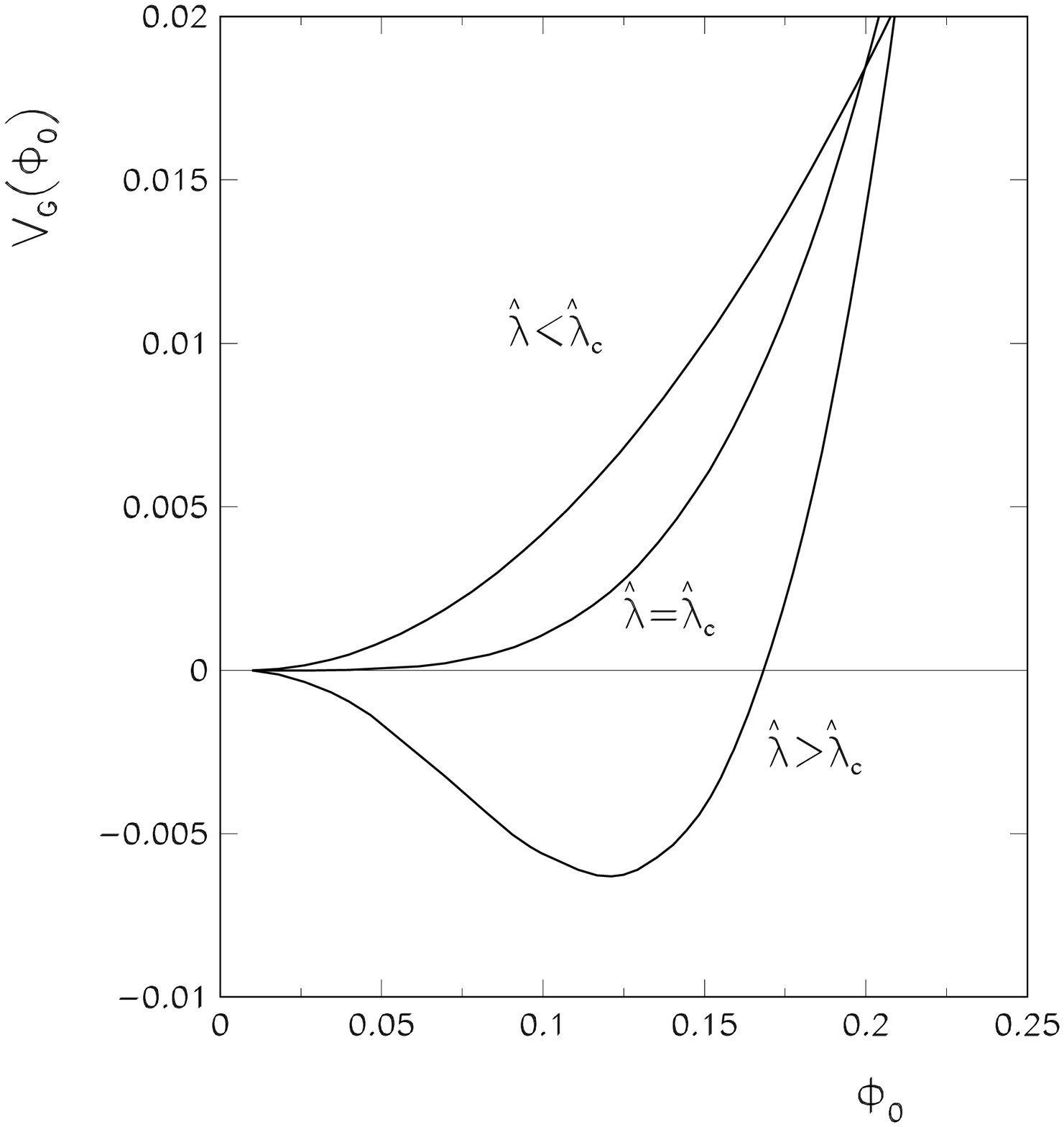}%
{FIG.~9        The second order generalized Gaussian effective potential for 
               $\nu=2$ and three different values of $\hat{\lambda}$.}
\InsertFigure{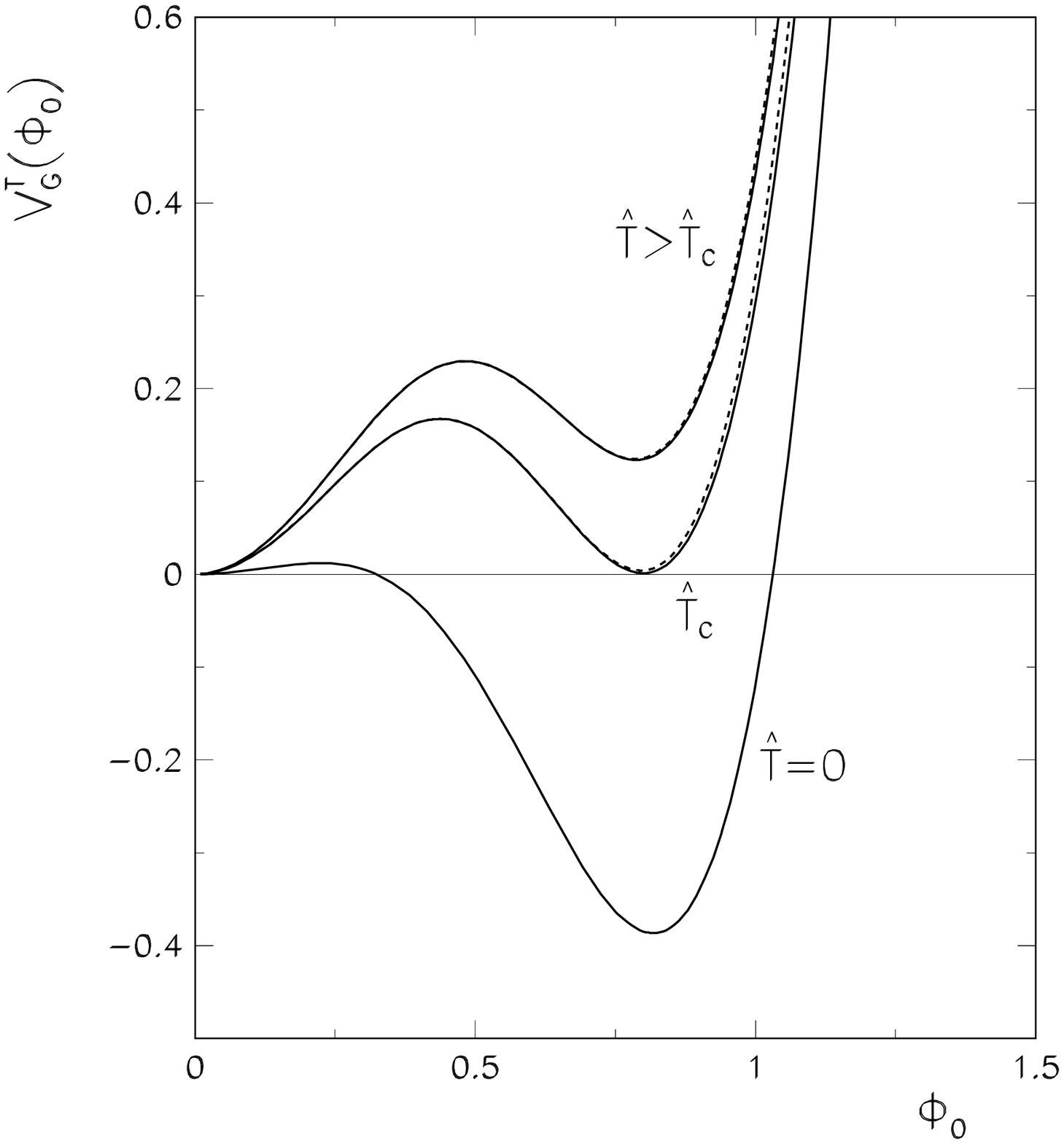}%
{FIG.~10       Lowest order thermal correction to the generalized Gaussian 
               effective potential for $\nu=1$ and $\hat{\lambda}=4$. Dashed
               lines refers to the high-temperature expansion.}
\InsertFigure{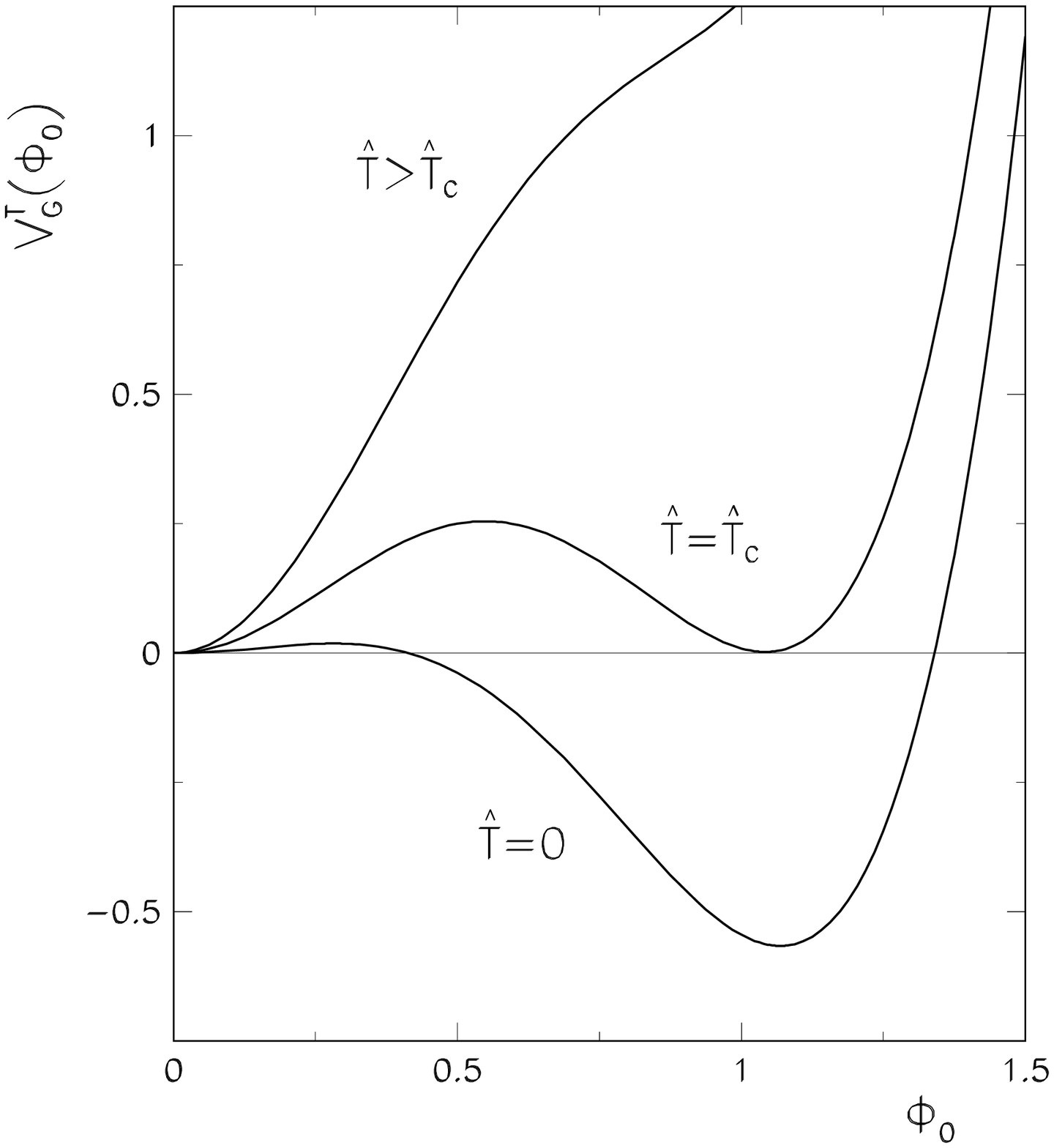}%
{FIG.~11       Lowest order thermal correction to the Gaussian effective 
               potential for $\nu=2$ and $\hat{\lambda}=4.$ The critical 
               temperature is $\hat{T}_c \simeq 1.60.$}
\InsertFigure{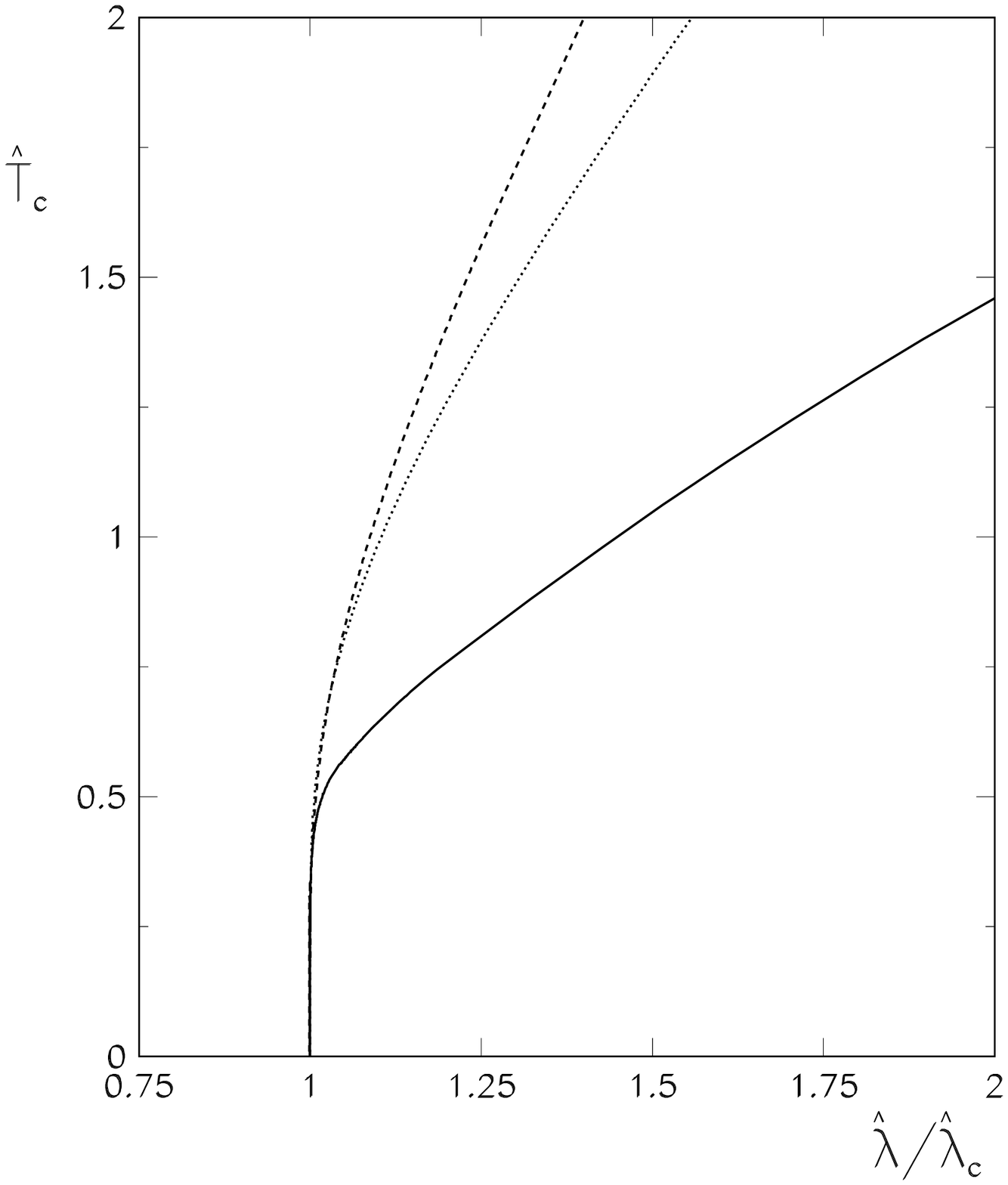}%
{FIG.~12       The critical temperature versus the coupling $\hat{\lambda}$
               for the one-loop effective potential (dotted line), the Gaussian 
               effective potential (dashed line), and the generalized Gaussian
               effective potential (full line) in two  spatial dimensions.}
\InsertFigure{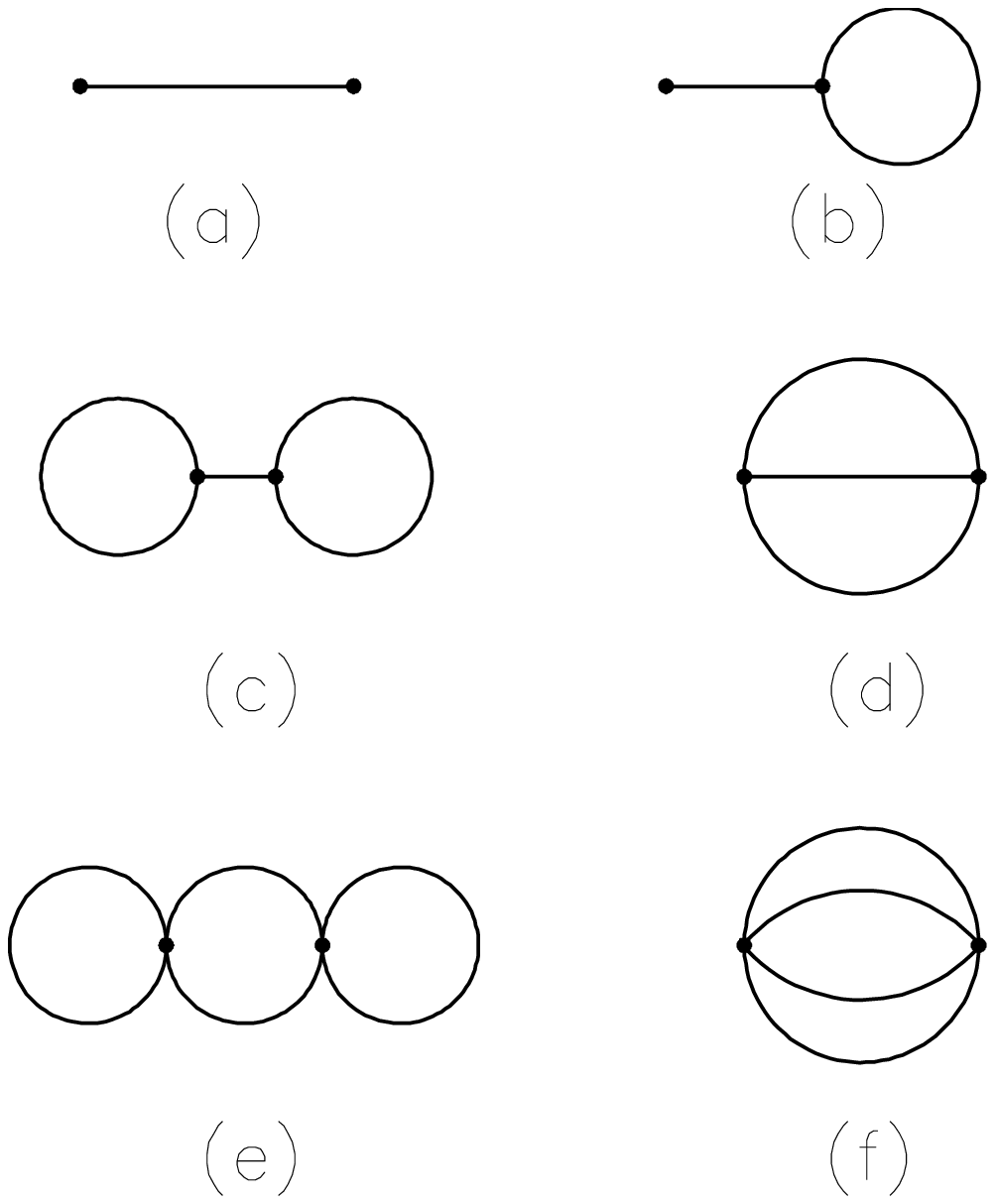}%
{FIG.~13       Thermal Feynman diagrams contributing to the second order 
               thermal corrections to the  generalied Gaussian effective 
               potential.}
\InsertFigureBis{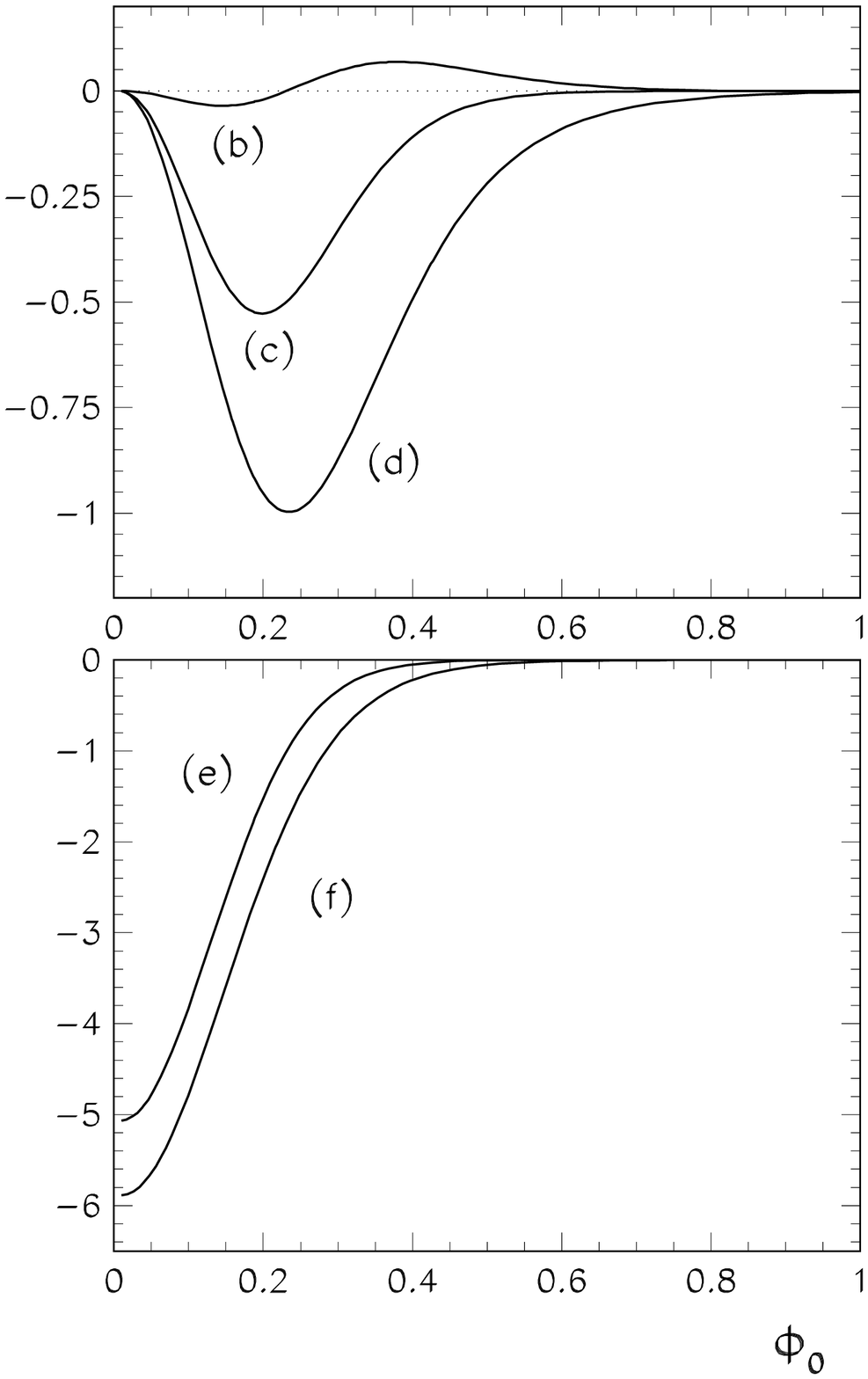}%
{FIG.~14       The second order thermal corrections to the generalized Gaussian 
               effective potential for $\nu=1$, $\hat{\lambda}=4$ and 
               $\hat{\beta}=1.$}
\InsertFigure{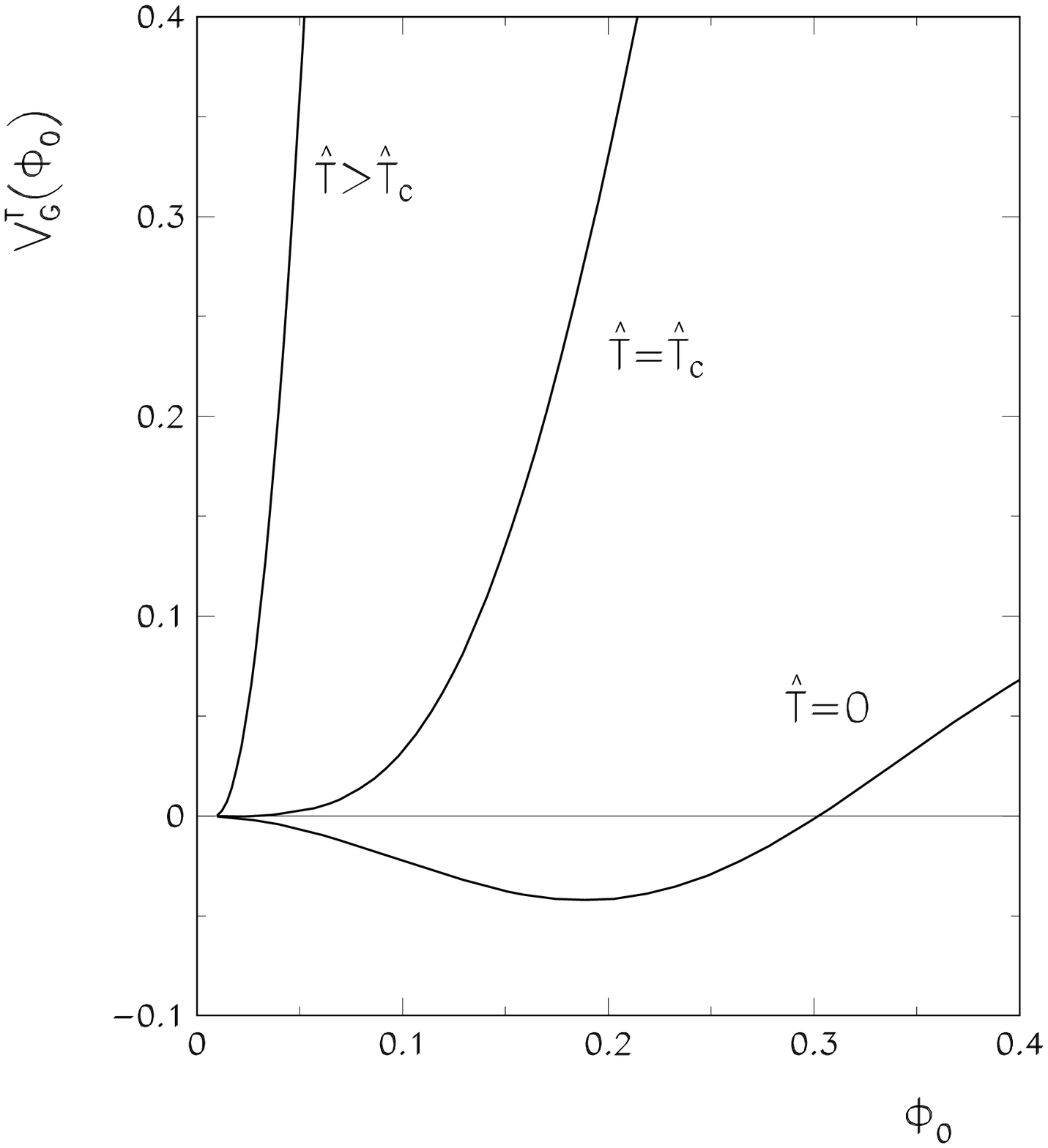}%
{FIG.~15       The generalized Gaussian effective potential with second order 
               thermal corrections $\nu=1$, $\hat{\lambda} =4$ and three 
               different values of the temperature.}
\InsertFigureBis{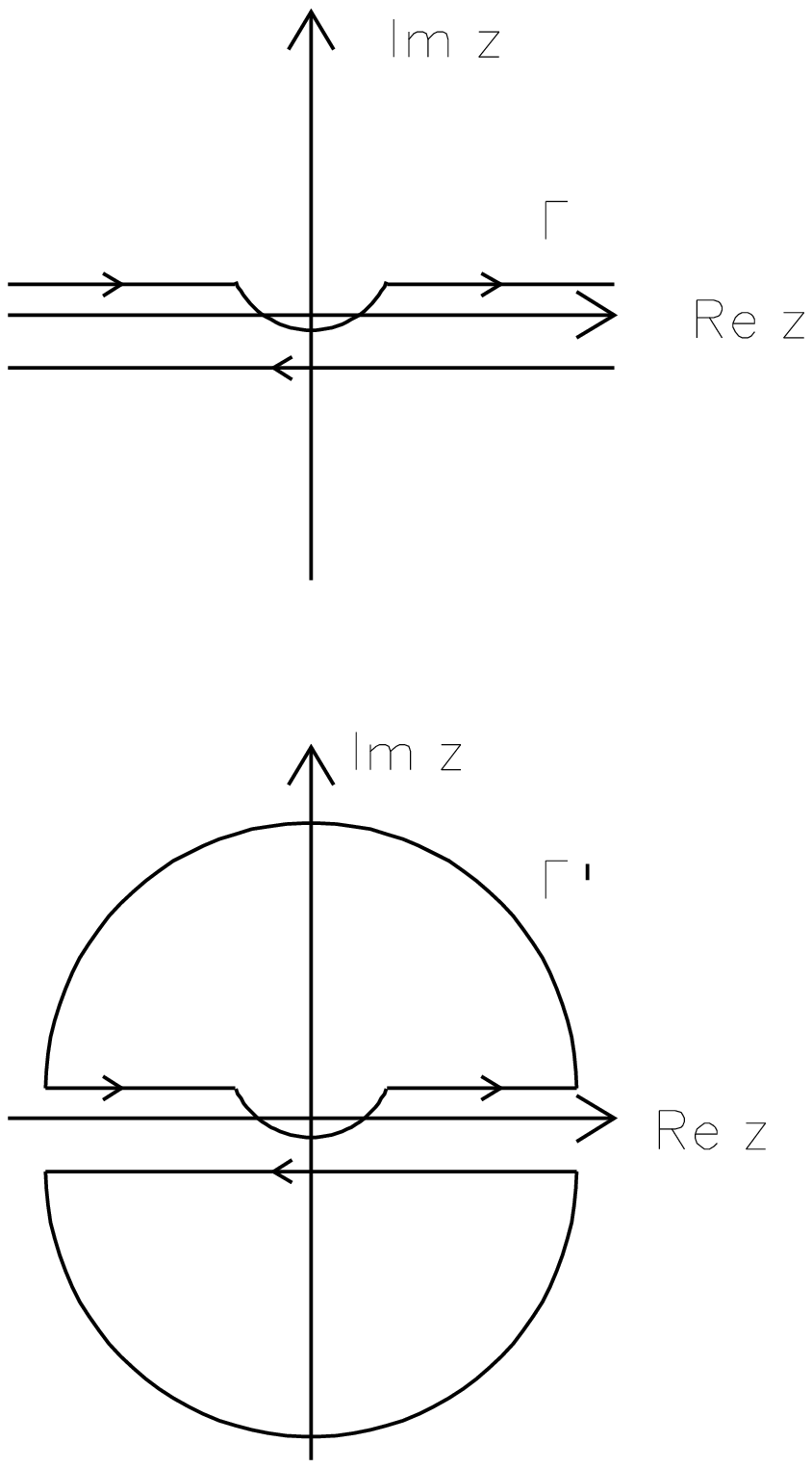}%
{FIG.~16       The contours $\Gamma$ and $\Gamma'$ in the complex z-plane.}

\begin{references}
%
\bibitem{Stev84}  P.M. Stevenson ,  Phys. Rev. {\bf D 30}, 1714 (1984); 
                  {\bf D 32}, 1389 (1985).
%
\bibitem{Feenberg69} For a review, see: E. Feenberg, {\it Theory of Quantum
                     Fluids} (Academic Press, New York, 1969); J.W. Clark, in
                     {\it The Many-Body Problem }, edited by R. Guardiola and
                     J. Ros, Lectures Notes in Physics, Vol. 138 (Springer,
                     Berlin, 1981); J.W. Clark and E. Krotscheck, in 
                     {\it Recent Progress in Many-Body Physics }, edited 
                     by H. K\"ummel and M. L. Ristig,
                     Lectures Notes in Physics, Vol. 198 (Springer,
                     Berlin, 1983).                      
%
\bibitem{S-R}      L. I. Schiff,  Phys. Rev. {\bf 130}, 458 (1963);
                   G. Rosen, Phys. Rev. {\bf 173}, 1632 (1968). 
%
\bibitem{C-J-T}    J. M. Cornwall, R. Jackiw, and E. Tomboulis,  Phys. Rev.
                   {\bf D 10}, 2428 (1974). 
%
\bibitem{Barnes}      T. Barnes and G.I. Ghandour,  Phys. Rev. {\bf D 22}, 924 
                   (1980).
%
\bibitem{Col}      S. Colemann and E. Weinberg, Phys. Rev. {\bf D 7}, 1888 
                   (1973);
                   S. Weinberg, Phys. Rev. {\bf D 7}, 2887 (1973);
                   R. Jackiw, Phys. Rev. {\bf D 9}, 1686 (1973).     
%
\bibitem{J-L}      G. Jona-Lasinio,  Nuovo Cimento {\bf 34}, 1790 (1964);
                   K. Symanzik,  Comm. Math. Phys. {\bf 16}, 48 (1970);
                   S. Coleman, "Secret Symmetry" in {\it Law of Hadronic 
                   Matter}, ed. A. Zichichi (Academic Press N.Y., 1975).
%
\bibitem{Cea90}    P. Cea, Phys. Lett. {\bf B 236}, 191 (1990).
%
\bibitem{Cea88}    A wider discussion can be found in: P. Cea, Phys. Rev.
                   {\bf D 37}, 1637 (1988).
%
\bibitem{B-G}      K. A. Brueckner,  Phys. Rev. {\bf 100}, 36 (1955);
                   J. Goldstone,  Proc. R. Soc. London {\bf A 239}, 
                   267 (1957).
%
\bibitem{F-W}      For a clear exposition, see: 
                   A.L. Fetter and  J.D. Walecka, {\it Quantum Theory of 
                   Many-Particle System} (Mc. Graw-Hill, N.Y. 1971).
%
\bibitem{G-M-L}    M. Gell-Mann and F. Low, Phys. Rev. {\bf 84}, 350 (1951).
%
\bibitem{C-T94}    P. Cea and  L. Tedesco,  Phys. Lett. {\bf B 335} 
                   423  (1994).
%
\bibitem{F88}      For a lucid discussion, see:
                   R. P. Feynman, {\it Variational Calculation in Quantum Field 
                   Theory}, L. Polley e D.E.L. Pottingen Editors (World 
                   Scientific 1988).
%
\bibitem{CH75}     S.J. Chang, Phys. Rev. {\bf D 12}, 1071 (1975);
                   Phys. Rep. {\bf C 23 }, 301 (1975).
%
\bibitem{CH76}     S.J. Chang, Phys. Rev. {\bf D 13}, 2778 (1976).
%
\bibitem{SG}       B. Simon and  R.G. Griffiths,  Comm. Math. Phys. 
                   {\bf 33}, 145 (1975).
%
\bibitem{RAMOND}   See, for instance: P. Ramond, {\it Field Theory a Modern 
                   Primer} (Addison Wesley, 1990).
%
\bibitem{P-R}      L. Polley and  U. Ritschel, Phys. Lett.
                   {\bf B 221}, 2778 (1989).
%
\bibitem{TH}       M. H. Thoma,  Zeit. Phys. {\bf C 44}, 343  (1989).
%
\bibitem{G-R}      I.S. Gradshteyn and I.M. Ryzhik, {\it Tables of Integrals, 
                   Series and Products} (Academic Press, 1980).
%
\bibitem{S-S}      I. Stancu and P.M. Stevenson, Phys. Rev. {\bf D 42}, 2710 
                   (1990).
%
\bibitem{S}        I. Stancu, Phys. Rev. {\bf D 43},  1283 (1991).
%
\bibitem{P-H-T}    A. Peter, J.M. H\"auser, M.H. Thoma, and W. Cassing, 
                   {\it Cluster
                   Expansion Approach to the Effective Potential in 
                   $\Phi^4_{2+1}$-Theory}, hep-th/9502103.
%
\bibitem{M}        S. F. Magruder,  Phys Rev. {\bf D 14}, 1602 (1976).
%
\bibitem{C-M}      S. Chang and S. F. Magruder, Phys. Rev. {\bf D 16}, 
                   983 (1977).
%
\bibitem{B}        C.W. Bernard,  Phys. Rev. {\bf D 9}, 3312 (1974).
%
\bibitem{W}        S. Weinberg, Phys. Rev. {\bf D 9}, 3357 (1974).
%
\bibitem{D-J}      L. Dolan and  R. Jackiw,  Phys. Rev.  {\bf D 9}, 3320 (1974).
%
\bibitem{H-S}      G.A. Hajj and  P.M. Stevenson, Phys. Rev. {\bf D 37}, 
                   413 (1988).
%
\bibitem{Rod}      I. Roditi, Phys. Lett. {\bf B 169 }, 264  (1986);
                   B. Alles and R. Tarrach, Phys. Rev. {\bf D 33}, 1718 (1986);
                   E. {\bf D 34},  664  (1986);
                   A. Bardeen and Moshe, Phys. Rev. {\bf D 34}, 1229 (1986);
                   A. Okopinska, Phys. Rev. {\bf D 35}, 1835 (1987); Phys. Rev.
                   {\bf D 36}, 2415 (1987).
%
\bibitem{MATSU}    T. Matsubara,  Prog. Theor. Phys. {\bf 14}, 351 (1955).
%
\bibitem{Feynman}  See, for instance: R. Feynman, {\it Statistical Mechanics} 
                   (W. Benjamin, 1982).
%
\bibitem{C-T2}     P. Cea and L. Tedesco, {\it Finite Temperature 
                   Generalized Gaussian Effective Potential}, Bari-Th 198/95.
%
\bibitem{A-G-D}    See, for instance: A. Abrikosov, L.P. Gorkov, and  I.E. 
                   Dzyaloshinki, {\it Methods of Quantum Field Theory in 
                   Statistical Physics} (ed. Dover publications N.Y. 1975).
%
\bibitem{G-R-H}    E. K. U. Gross, E. Runge, and O. Heinonen, {\it 
                   Many-Particle Theory} (Adam Hilger, Bristol, 1991).
%
\bibitem{K-M}      M.B. Kislinger and P.D. Morley, Phys. Rev. {\bf D 13}, 2779 
                   (1976).
%
\bibitem{C-T3}     P. Cea and L. Tedesco, {\it Generalized Gaussian Effective 
                   Potential: Second Order Thermal Corrections}, Bari-Th
                   208/95. 
%
\bibitem{N-G-P}    H. W. L Naus, T Gasenzer, and H.J. Pirner, {\it Effective 
                   Hamiltonian  for Scalar Theories in the Gaussian 
                   Approximation}, hep-ph/9507357.
%
\bibitem{F-F-S}    R. Fernandez, J. Fr\"ohlich, and A. D. Sokal, {\it 
                   Random Walks, Critical Phenomena, and Quantum Field 
                   Theory} (Springer-Verlag, Berlin, 1992).
%
\bibitem{Consoli}  M. Consoli and P.M. Stevenson, Zeit. Phys. {\bf C 63},
                   427 (1994).
%
\bibitem{A-A-C}    A. Agodi, G. Adronico, and  M. Consoli,  Zeit. Phys. 
                   {\bf  C 66}, 439 (1995).
%
\bibitem{C-C-C-F}  P. Cea, L. Cosmai, M. Consoli, and R. Fiore, {\it 
                   Lattice effective potential of $(\lambda \Phi^4)_4$:
                   nature of the phase transition and bounds on the Higgs
                   mass}, hep-th/9603019.
%
\bibitem{Somm}     A. Sommerfeld, {\it Partial Differential Equations in 
                   Physics} (Academic Press, N.Y. 1967).
%
\end{references}
\end{document}